\documentclass[iop,apj,twocolappendix]{emulateapj}
\usepackage{amsmath,amssymb}
\usepackage{graphicx}
\usepackage{epstopdf}
\usepackage{natbib}
\usepackage{color}
\usepackage[breaklinks,colorlinks=true,urlcolor=blue,linkcolor=red,citecolor=blue]{hyperref}
\graphicspath{{./}{figs/}}

\newcommand{\angstrom}{{\rm \mathring A}}
\begin{document}

\title{A surprising excess of radio emission in extremely stable quasars: a unique clue to jet launching?}
\author{Wen-Yong Kang$^*$\altaffilmark{1,2}, Jun-Xian Wang$^*$\altaffilmark{1,2}, Zhen-Yi Cai$^*$\altaffilmark{1,2}, Hao-Chen Wang\altaffilmark{1,2}, Wen-Ke Ren\altaffilmark{1,2}, Mai Liao\altaffilmark{3,4,5},  \\
Feng Yuan\altaffilmark{6},
Andrzej Zdziarski\altaffilmark{7},
Xinwu Cao\altaffilmark{8}}
\email{kwy0719@ustc.edu.cn, jxw@ustc.edu.cn, zcai@ustc.edu.cn}
\affil{
$^1$CAS Key Laboratory for Research in Galaxies and Cosmology, Department of Astronomy, University of Science and Technology of China, Hefei 230026, China;\\
$^2$School of Astronomy and Space Science, University of Science and Technology of China, Hefei 230026, China;\\
$^3$National Astronomical Observatories, Chinese Academy of Sciences, 20A Datun Road, Chaoyang District, Beijing 100101, China;\\
$^4$Chinese Academy of Sciences South America Center for Astronomy, National Astronomical Observatories, CAS, Beijing 100101, China;\\
$^5$Instituto de Estudios, Astrofísicos Facultad de Ingeniería y Ciencias, Universidad Diego Portales, Av. Ejército 441, Santiago, Chile;\\
$^6$Center for Astronomy and Astrophysics and Department of Physics, Fudan University, 2005 Songhu Road, Shanghai 200438, China;\\
$^7$Nicolaus Copernicus Astronomical Center, Polish Academy of Sciences, Bartycka 18, PL-00-716 Warszawa, Poland\\
$^8$Institute for Astronomy, School of Physics, Zhejiang University, 866 Yuhangtang Road, Hangzhou 310058, China
}

\begin{abstract}

Quasars are generally divided into jetted radio-loud and non-jetted radio-quiet ones, but why only 10\% quasars are radio loud has been puzzling for decades. Other than jet-induced-phenomena, black hole mass, or Eddington ratio, prominent difference between jetted and non-jetted quasars has scarcely been detected. Here we show a unique distinction between them and the mystery of jet launching could be disclosed by a prominent excess of radio emission in extremely stable quasars (ESQs, i.e., type 1 quasars with extremely weak variability in UV/optical over 10 years). Specifically, we find that $>$ 25\% of the ESQs are detected by the FIRST/VLASS radio survey, while only $\sim$ 6-8\% of the control sample, matched in redshift, luminosity, and Eddington ratio, are radio-detected. The excess of radio detection in ESQs has a significance of 4.4 $\sigma$ (99.9995\%), and dominantly occurs at intermediate radio loudness with R $\sim$ 10 -- 60. The radio detection fraction of ESQs also tends to increase in the ESQ samples selected with more stringent thresholds. Our results are in contrast to the common view that RL quasars are likely more variable in UV/optical due to jet contribution. New clues/challenge posed by our findings highlight the importance of extensive follow-up observations to probe the nature of jets in ESQs, and theoretical studies on the link between jet launching and ESQs.  Moreover, our results makes ESQs, an essential population which has never been explored, unique targets in the burgeoning era of time domain astronomy, like their opposite counterparts of quasars exhibiting extreme variability or changing-look features.

\end{abstract}

\section{Introduction}
\label{introduction}

Based on the relative radio intensity, i.e. radio loudness, which is defined as the ratio of the radio flux density to the optical one \citep[e.g.][]{1989AJ.....98.1195K}, quasars could be divided into radio-loud (RL) and radio-quiet (RQ) ones, or jetted and non-jetted ones \citep{2017A&ARv..25....2P, 2019NatAs...3..387P}. 
The RL quasars, with powerful relativistic and well collimated jets \citep{2019ARA&A..57..467B}, are typically 1000 times brighter in radio than the RQ ones \citep{1993MNRAS.263..425M, 2019NatAs...3..387P}. While it is generally believed that the rotational energies of the black hole (BH) or the inner accretion flow could be extracted to power jets \citep{BZ1977, BP1982, 2019ARA&A..57..467B},
why the relativistic jets have only been launched in a small population ($\sim$ 10\%) of quasars has been a mystery for many decades. This is particularly puzzling considering that both the RL and RQ quasars have rather similar, except in radio, spectral energy distributions \citep[e.g.][]{Elvis1994,Shang2011}, suggesting both populations are powered by a similar accretion process. Furthermore, \cite{Dunlop2003, 2004MNRAS.355..196F, 2013MNRAS.429....2F} found that hosts of both RL and RQ quasars are similarly massive galaxies, and they remain star forming \citep{2013MNRAS.429....2F, 2016ApJ...831..168K}, suggesting the host does not dominantly determine the radio loudness of quasars. While observations have revealed that the RL quasars have more massive BHs and smaller Eddington ratios (but with strong overlap in range) compared with the RQ ones \citep[e.g.][]{Laor2000, Lacy2001, Ho2002}, searching for observational differences (other than jet induced phenomena) between RL and RQ quasars with matched BH mass and Eddington ratio may yield essential clues to understanding the jet launching mystery.

Variability has been a defining characteristic of AGNs and quasars \citep[e.g.][]{Ulrich1997}. Studying the variability, particularly in UV/optical which is believed to be predominantly driven by magnetic turbulence in the accretion disc, can be of great help in probing the yet unclear underlying physics of the inner accretion process.  
Plentiful observational studies have reported correlations between the UV/optical variation and several known parameters, including the luminosity, rest frame wavelength (i.e., stronger variability at shorter wavelengths), Eddington ratio, BH mass, redshift \citep{berk2004ensemble, Wilhite2005, Wold2007, wilhite2008variability, Bauer2009, 2010ApJ...716L..31A, Macleod2010, 2011A&A...525A..37M, Zuo2012, 2013A&A...560A.104M, Kozlowski2016ApJ826, 2018ApJ...866...74S}, and new parameters including  X-ray loudness and equivalent width of emission lines  \citep{2018ApJ...868...58K, 2021ApJ...911..148K}. 
As for the RL quasars, they often exhibit stronger rapid (e.g., intra-night) variability \citep[e.g.][]{Gupta2005} and marginally stronger long-term variability \citep[e.g.][]{helfand2001long,berk2004ensemble}, compared with the RQ ones, which could be attributed to the contribution of the UV/optical emission from the jet which could be more variable than the disk component.

In the era of time domain astronomy, great attention of the community has been attracted to AGNs exhibiting violent variability, e.g., changing-look (CL) AGNs and quasars \citep[e.g.][]{Cohen1986, LaMassa2015, Macleod2016, Sheng2017, Yang2018,Sheng2020ApJ...889...46S,  Green2022, Ricci2022}, and extremely variable quasars \citep[EVQs, ][]{2018ApJ...854..160R, 2019ApJ...874....8M, 2022ApJ...925...50R}, which are a small population of sources showing strong variability likely driven by yet unclear violent changes in the inner accretion disc activity. 
In contrast, in this work we focus on extremely stable quasars (ESQs), which exhibit rather weak or undetectable long-term (over 10 years) variability in UV/optical. Such quasars, the opposite counterparts of CL quasars and EVQs in the parameter space of variability, have never been studied in literature. We find that ESQs exhibit significant excess of radio emission compared with the normal quasars, providing unique new clues to jet launching in quasars.
Throughout this work, cosmological parameters of $H_0=70\rm\ km~s^{-1}~Mpc^{-1}$, $\Omega_{\rm m}=0.3$, and $\Omega_{\Lambda}=0.7$ are adopted. 
 
\section{Selection of extremely stable quasars}
\label{S:sample}

We adopt the 10-year-long light curves \citep{Macleod2012} for the 9258 spectroscopically confirmed quasars in the Sloan Digital Sky Survey (SDSS) Stripe 82, which has been scanned around 60 times in the $ugriz$ bands \citep{Sesar2007}, to calculate the long-term variability amplitude of quasars. We do not include additional photometric data of them from other time domain surveys, such as Pan-STARRS1 \citep{Flewellin2020}, because the different filter transmissions between surveys would cause extra scatter to the variability measurements. 
As the SDSS $g$ and $r$ bands have the best photometry for quasars in Stripe 82 \citep[see Fig. 2 of][]{Sun2014}, and the intrinsic variability of quasars is expected to be stronger at shorter wavelength, in this work we primarily adopt $g$ band to select ESQs, and utilize the other bands to secure the selection.
We focus on 9146 quasars from \cite{Macleod2012} which have at least 10 photometric measurements in the $g$-band light curve, after excluding a few unphysical data points that may be present. In addition, to use the up-to-date physical properties, such as BH mass and Eddington ratio, for these quasars, we also drop a few sources for which no counterpart in the SDSS data release 16 quasar (DR16Q) catalogue is found within 2 arcsec or the matched counterpart does not have a valid measurement on the BH mass by \citet{Wu2022}.

The excess variance, $\sigma_{\mathrm{rms}}^2$, has widely been utilized to quantify the quasar variability with a single model-independent parameter \citep[e.g.][]{2003MNRAS.345.1271V}.
However, as demonstrated in Appendix \ref{appendixA} for the ESQs, the canonical $\sigma_{\mathrm{rms}}^2$ could be significantly biased by a minority of epochs with too large photometric uncertainties. We thus revise the canonical calculation of $\sigma_{\mathrm{rms}}^2$ 
by adding weight to each photometric measurement (see Appendix \ref{appendixA}).

\begin{figure}
  \centering
  \includegraphics[width=\linewidth]{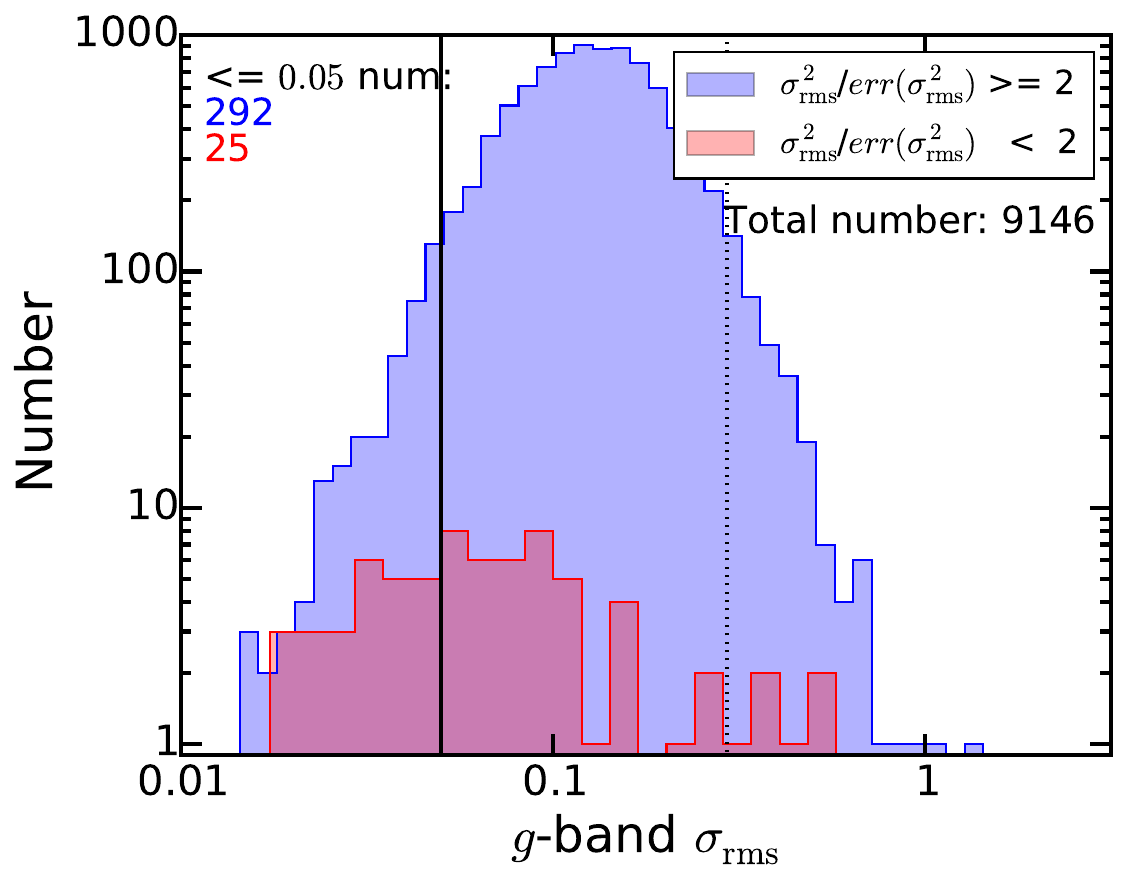}
  \caption{Distribution of the $g$-band $\sigma_{\mathrm{rms}}$ of 9146 SDSS quasars in Stripe 82. For sources whose $\sigma_{\mathrm{rms}}^2$ are statistically non-detected with S/N($\sigma_{\mathrm{rms}}^2$) $<$ 2, upper limits to $\sigma_{\mathrm{rms}}$ are assigned (red).
  To the left of the vertical solid black line  ($\sigma_{\mathrm{rms}}$ = 0.05), there are 317 quasars with smallest $g$-band variability in the sample. For comparison, the vertical dashed line marks (to its right) equal number of sources with the strongest $g$-band variability. 
}
  \label{Fig2_2_1}
\end{figure}

\begin{figure}
  \centering
  \includegraphics[width=\linewidth]{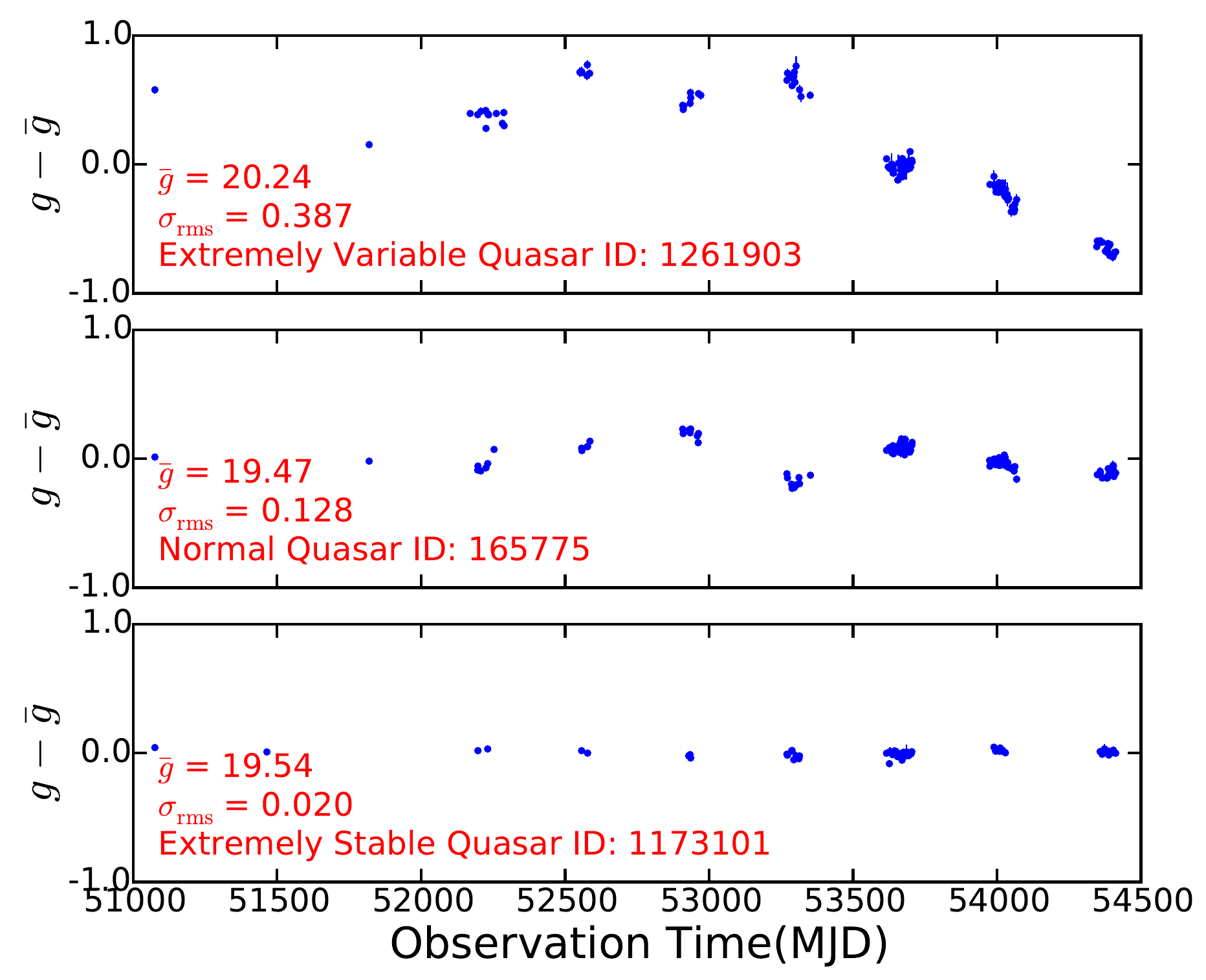} 
  \caption{An illustration of the $g$-band light curves of an EVQ (maximal $\Delta g > 1$ mag; top panel), a normal quasar (median panel), and an ESQ (bottom panel). The quasar IDs marked in plot are from \cite{Macleod2012}.
}
  \label{Fig2_2_2}
\end{figure}

We calculate $\sigma_{\mathrm{rms}}^2$ and \emph{error}($\sigma_{\mathrm{rms}}^2$) 
for each quasar in all SDSS bands. For sources with variability statistically non-detected in one band, i.e., with S/N($\sigma_{\mathrm{rms}}^2$) = $\sigma_{\mathrm{rms}}^2/\emph{error}(\sigma_{\mathrm{rms}}^2) < 2$, we adopt $2\times$\emph{error}($\sigma_{\mathrm{rms}}^2$) as the upper limit to $\sigma_{\mathrm{rms}}^2$. 
Distribution of the derived $\sigma_{\mathrm{rms}}$ in $g$ band is illustrated in Fig. \ref{Fig2_2_1}, where we can see that the ESQs are the opposite counterparts of the EVQs in the parameter space of variability amplitude.

We consider a series of thresholds of $\sigma_{\mathrm{rms}}$ to define ESQs. Any source would be identified as ESQs if

1. the $g$-band $\sigma_{\mathrm{rms}}$ is less than the threshold, and

2. the $u$-, $r$-, $i$-, and $z$-band $\sigma_{\mathrm{rms}}$ are all either less than the threshold or non-detected.

Using thresholds of $\sigma_{\mathrm{rms}}$ $\leq$ 0.02, 0.03, 0.04, and 0.05 magnitudes, we identify 3, 25, 53, and 136 ESQs, respectively.  In Fig. \ref{Fig2_2_2} we illustrate a typical $g$-band light curve of an ESQ, in comparison with a normal quasar and an EVQ.  Note the SDSS provides PSF magnitude for quasars which is unbiased and optimal for point-like sources \citep{Stoughton2002}. The host contamination is expected to be weak for luminous quasars, especially for our ESQs which tend to have even higher luminosities compared with all SDSS quasars (see Fig. \ref{L_z}). 

\section{Excess of radio emission in ESQs}
\label{S:radio}

\begin{figure}
  \centering
  \includegraphics[width=\linewidth]{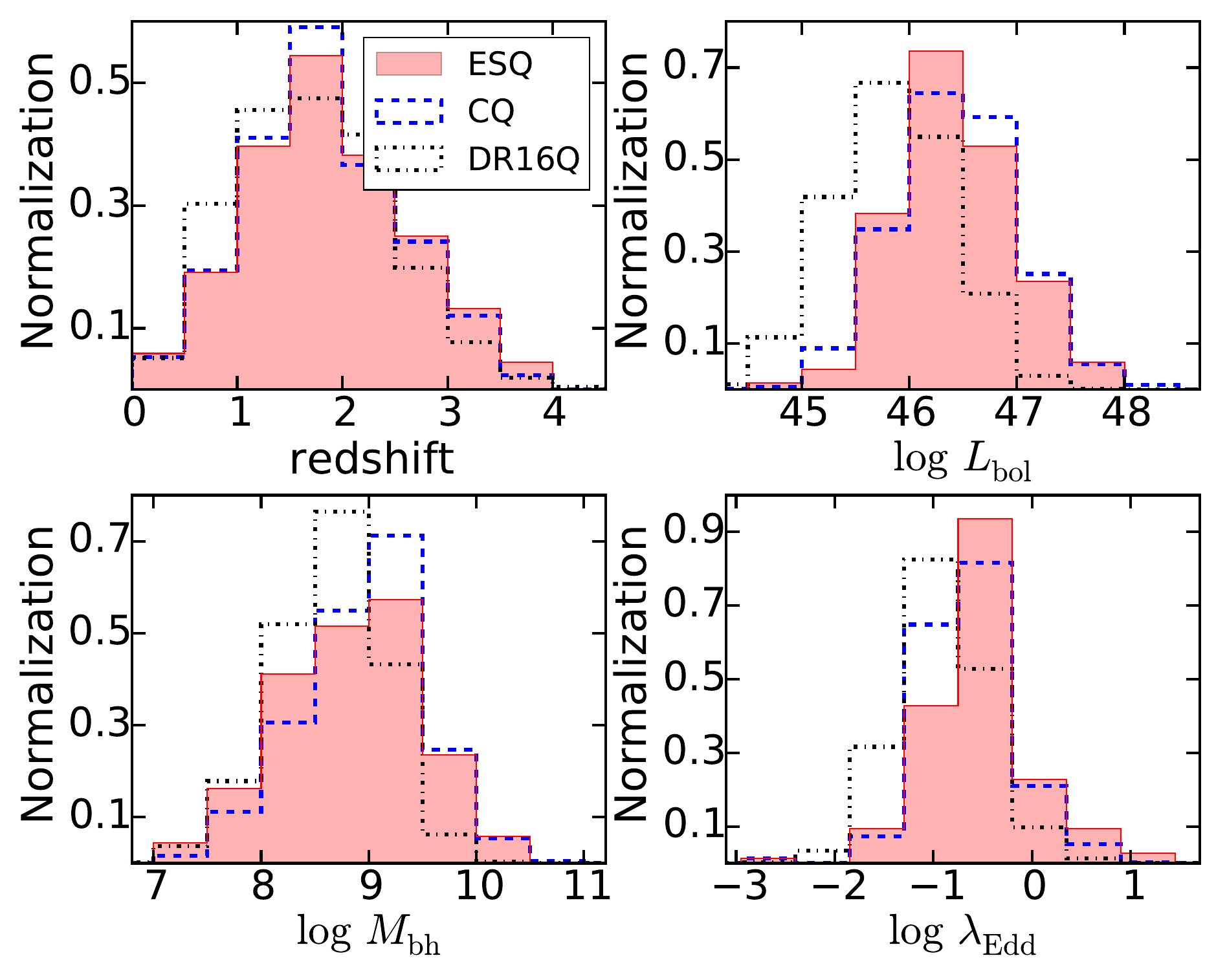}
  \caption{Normalized distributions of redshift, bolometric luminosity, BH mass, and Eddington ratio of all ESQs  (red) selected with threshold of $\sigma_{\mathrm{rms}} \leq 0.05$, compared to those of  the control quasars (CQ, blue) and all quasars in the SDSS DR16Q with valid measurements (black). All values are taken from \citet{Wu2022}.
}
  \label{L_z}
\end{figure}  

\begin{figure}
  \centering
  \includegraphics[width=\linewidth]{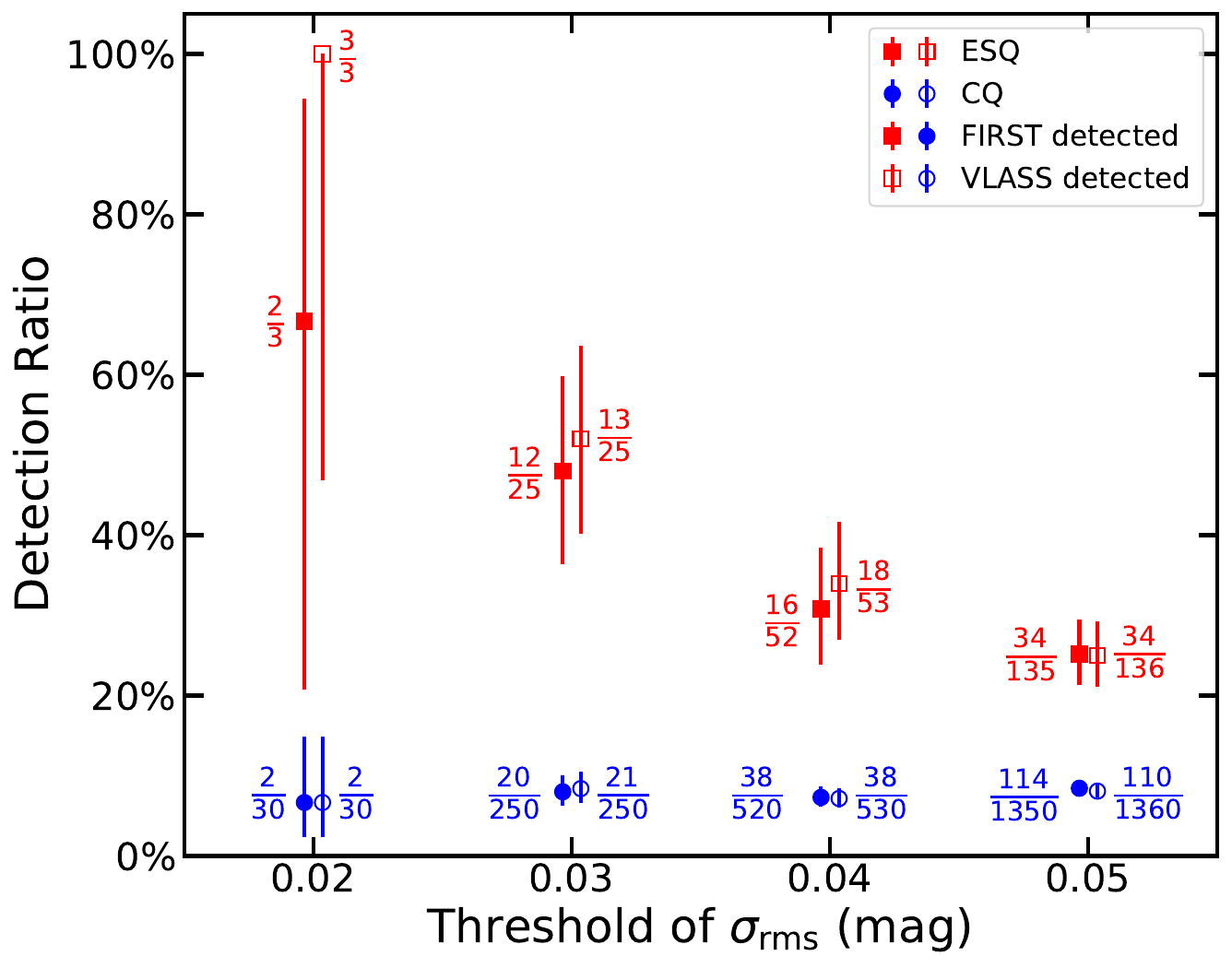}
  \caption{The radio (FIRST/VLASS) detection fractions of ESQs (filled or open squares) selected with a series of thresholds of $\sigma_{\mathrm{rms}}$, compared with control samples of quasars (CQs; filled or open circles), matched in redshift, magnitude, and BH mass.  
  Numbers of ESQs and CQs (denominator) and of radio-detected ones (numerator) are marked. 
  Error bars are  $1\sigma$ confidence limits based on a combination of Poisson and binomial statistics \citep[][]{Gehrels1986}. See Section~\ref{S:sample} and Appendix \ref{appendixB} for details of selections of the ESQs and control samples, respectively.
}
  \label{Fig3_1_1}
\end{figure}

As shown in Fig. \ref{L_z}, the selected ESQs span broad ranges of redshift, luminosity, BH mass, and Eddington ratio. To erase potential observational biases, we build control samples of quasars (CQs), with redshift ($z$), $g$-band apparent magnitude, and BH mass ($M_{\mathrm{bh}}$), matched to our ESQs. Note the simultaneous match of redshift, apparent magnitude, and BH mass automatically matches the luminosity and Eddington ratio. See Appendix \ref{appendixB} for details of the control sample selection.

\begin{figure}
  \centering
  \includegraphics[width=\linewidth]{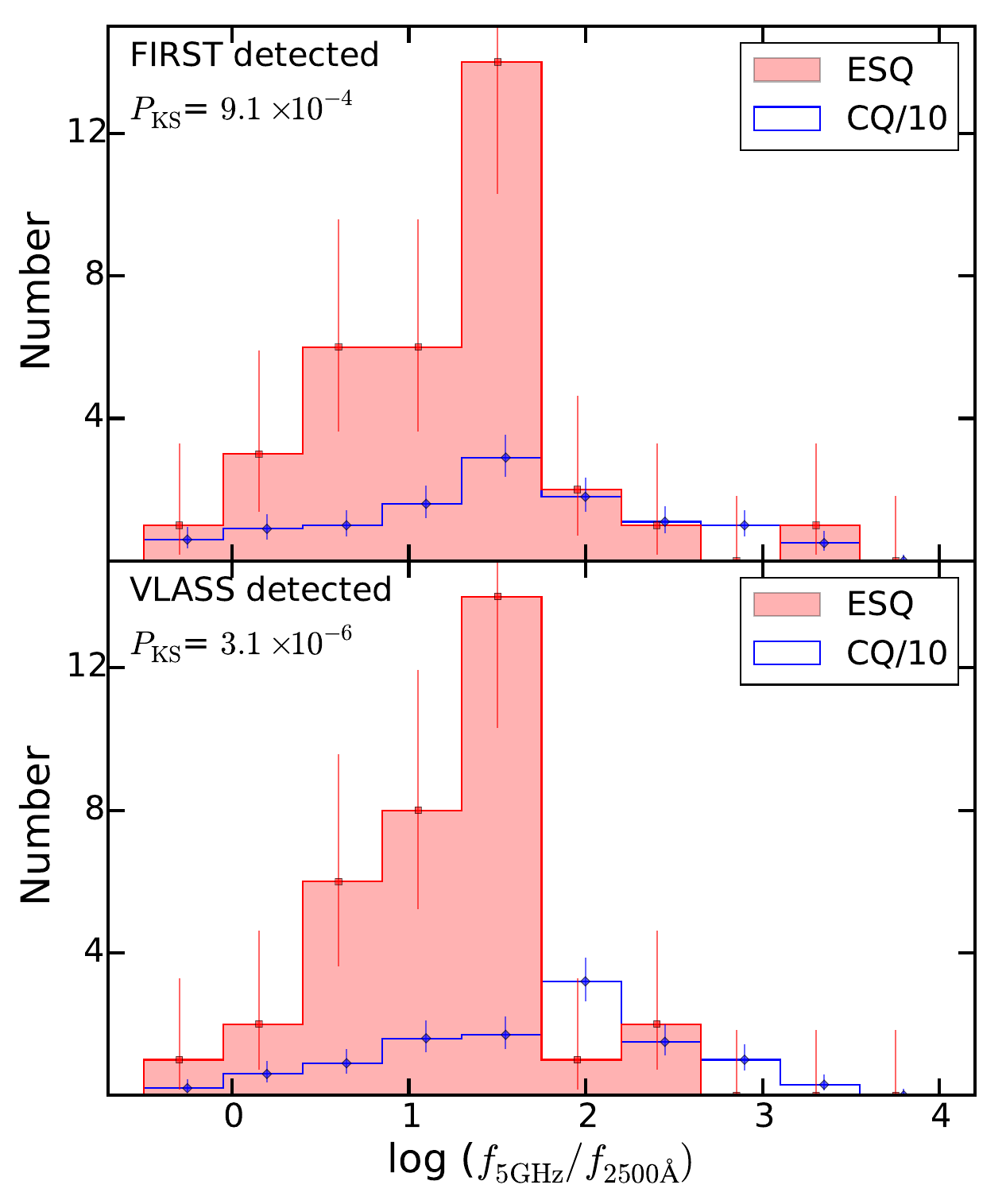}
  \caption{Distributions of the radio loudness of the radio-detected (upper: FIRST detected; lower: VLASS detected) ESQs and CQs.  Error bars are 1$\sigma$ confidence limits for Poisson statistics. The $P_{\mathrm{KS}}$ is the p-value of the Kolmogorov-Smirnov test between ESQs and CQs.
}
  \label{Fig3_3_1}
\end{figure}

We match our ESQ and the control samples to the FIRST \citep[1.4 GHz, ][]{1997ApJ...475..479W} and VLASS epoch 1 \citep[3 GHz, ][]{Gordon2021} catalogs, with a matching radius of 5\arcsec\ (see Appendix \ref{appendixD}). The matched results are displayed in Fig. \ref{Fig3_1_1}. The radio detection fraction of ESQs is remarkably high, $>$ 25\%, dependent on the $\sigma_{\mathrm{rms}}$ threshold adopted to identify ESQs. 
Moreover, the FIRST/VLASS detection fractions of the ESQs are significantly higher than that of the control samples ($\sim$ 6.7\% -- 8.4\%), and the difference gradually decreases towards higher threshold of $\sigma_{\mathrm{rms}}$. This indicates that the more stable an ESQ is, the more likely it would be detected in radio. The clear excess of radio emission in ESQs is further confirmed with $\sigma_{\mathrm{rms}}$ of SDSS quasars measured with Pan-STARRS1 \citep{Flewellin2020} and Gaia DR3 \citep{GaiaDR3} time domain photometry (W. K. Kang et al. 2024 and H. C. Wang et al. 2024, in preparations). In this work we focus on ESQs selected using the SDSS Stripe 82 light curves, which have $\sim$ 10 years time span with $\sim$ 60 visits per band, and are significantly longer/more than those of Pan-STARRS1 and Gaia, thus the variability could be more stringently constrained for ESQs. 

We further confirm that the ESQs we selected are type~1 quasars with at least one broad emission line significantly detected, and the dominant fraction of them have UV-to-optical color consistent with their control quasars (see Appendix \ref{appendixC}). These factors, together with their high luminosity and Eddington ratio (see Fig. \ref{L_z}),  indicate their extremely low variability is not dominated by strong host contamination.
While it is worthy of further exploring whether redder colors of a small fraction of ESQs are intrinsic or due to obscuration,  
excluding these redder ESQs does not alter our results (see Appendix \ref{appendixC}).  

\begin{figure*}
  \centering
  \includegraphics[width=12cm]{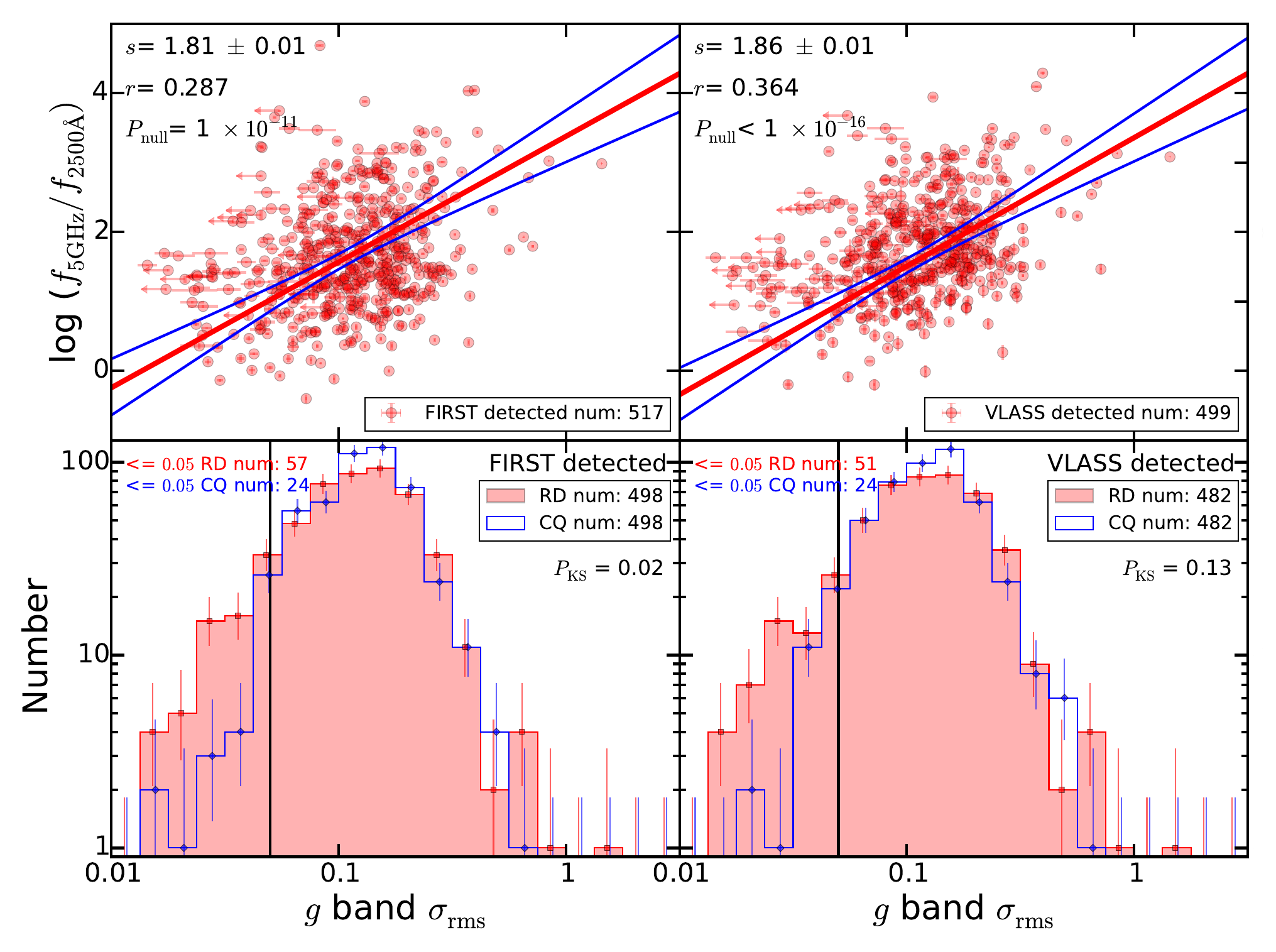}
  \caption{Upper: correlation between radio loudness $R$ and $g$-band $\sigma_{\mathrm{rms}}$ for the radio-detected (left: FIRST detected; right: VLASS detected) quasars in the SDSS Stripe 82. The  red and blue lines represent the best-fit bisector linear regression \citep{1990ApJ...364..104I} and the corresponding 3$\sigma$ confidence interval, respectively.  Shown in the upper-left corners are the bisector regression slope ($s$ with the corresponding 1$\sigma$ error), the Pearson's rank correlation coefficient ($r$), and the significance level ($P_{\mathrm{null}}$). Numbers of radio-detected quasars are shown in the lower-right corners. Upper limits of $\sigma_{\mathrm{rms}}$ for minor quasars with ${\rm S/N}(\sigma^2_{\mathrm{rms}}) < 2$ are indicated by the leftward arrows.
  Lower: histogram distributions of the $g$-band $\sigma_{\mathrm{rms}}$ for the radio-detected quasars (RD) in the SDSS Stripe 82, compared with the corresponding control samples of quasars (CQ). Error bars to the histogram are 1$\sigma$ confidence limits for Poisson statistics. Shown in the upper-right corners are numbers of radio-detected quasars with matched control ones (These numbers are different from ones in upper panels because part of radio-detected quasars can not find control quasar). The $P_{\mathrm{KS}}$ is the p-value of the Kolmogorov-Smirnov test between the radio-detected sample and the control sample.
}
  \label{Fig3_5_1}
\end{figure*}

\section{Discussion}
\label{S:discussion}

\begin{figure*}
  \centering
  \includegraphics[width=12.5cm]{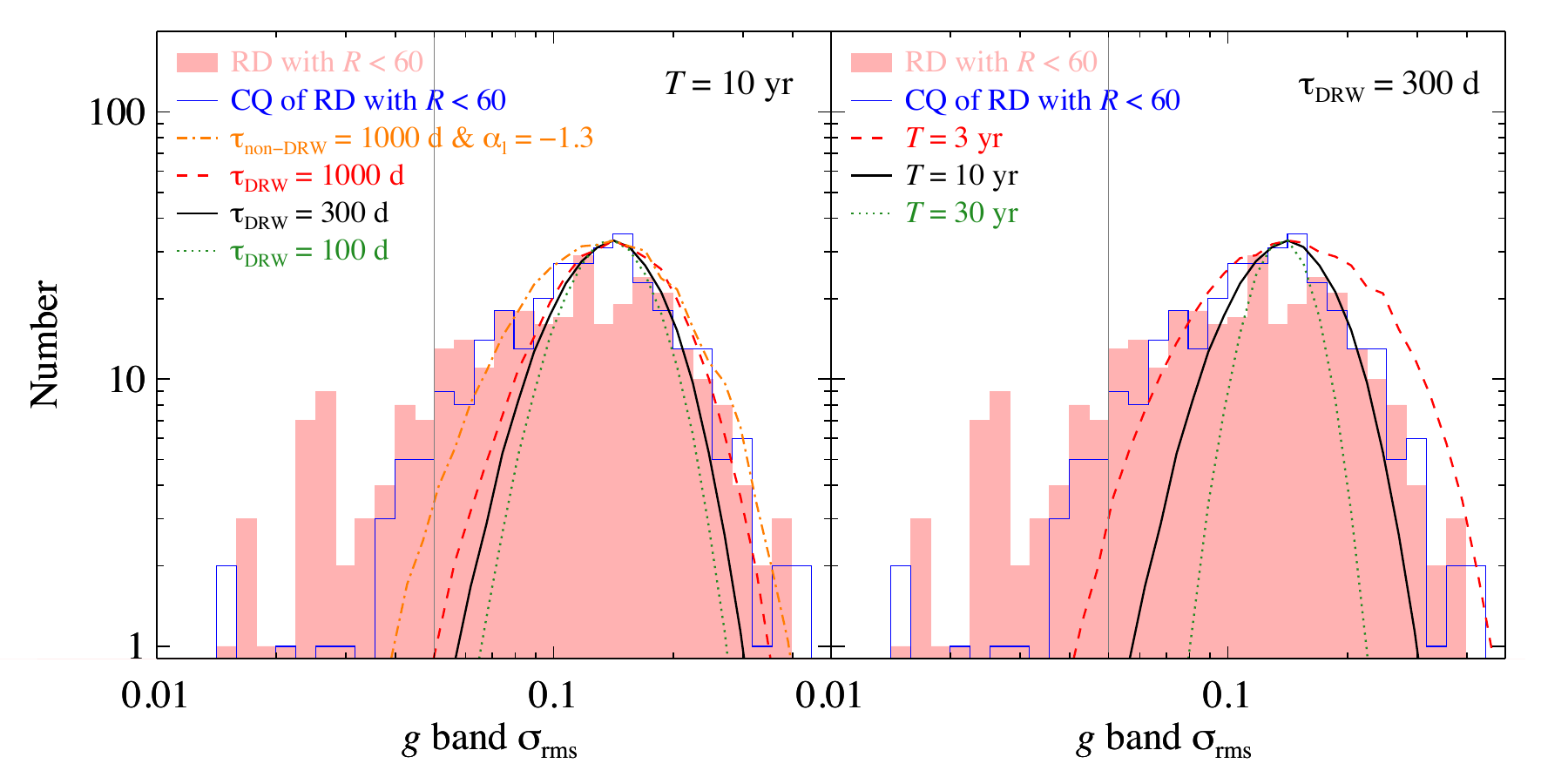}
  \caption{ Left: similar to the lower left panel of Fig. \ref{Fig3_5_1}, but here we only plot the $g$-band $\sigma_{\mathrm{rms}}$ distribution for the FIRST detected quasars (RD) with $R < 60$ and the corresponding control sample (CQ). We omit the Poisson error bars to avoid confusion. We then plot the expected distribution for a single normal variable quasar using mock light curves, considering the effect of stochastic nature of variation, limited length of light curve, and sparse sampling (see Appendix \ref{appendixE}). 
  Right: similar to the left panel, we here plot the simulated distribution illustrating the effect of the various lengths of the mock light curves (3 years, 10 years, and 30 years). 
}
  \label{fig-simulation}
\end{figure*}

As aforementioned in \S\ref{introduction}, the RL quasars are known to show (marginally) stronger variability in UV/optical compared with the RQ ones, likely due to the jet contribution. Contrarily, the prominent radio excess in ESQs reported here is indeed out of expectation. To determine the radio physical properties of these ESQs and the relationship to the general quasars, we adopt the canonical definition of the radio loudness, $R = f_{\rm 5GHz}/f_{2500\angstrom}$, where $f_{2500\angstrom}$ is the rest-frame 2500$\angstrom$ flux density from \cite{Wu2022}, and $f_{\rm 5GHz}$ is the 5 GHz flux density inferred from the observed-frame 1.4 GHz (FIRST) or 3 GHz (VLASS) flux densities assuming $f_{\nu} \propto {\nu}^{-\alpha}$ with a radio spectral index of $\alpha$ = 0.5 \citep[][see also Appendix \ref{appendixD}]{Jiang2007ApJ...656..680J}. Fig. \ref{Fig3_3_1}  plots distributions of the radio loudness for the radio-detected ESQs and CQs. 
We see the excess of the radio-detected ESQs 
mainly occurs in the radio intermediate regime ($R \sim$ 10 -- 60) which
clearly requires jet. The excess is also visible at $R \sim$ 1 -- 10 which falls in the radio quiet regime (likely due to weaker jet in those sources). However, the excess shows a sharp cutoff and disappears in the radio loud regime ($R > 60$). While we see no statistical difference between ESQs and CQs at $R > 60$ (partially due to the small sample size), the lack of excess ESQs at $R > 60$ is statistically significant (compared with that at $R < 60$, with a p-value of $\sim$ 0.001), probably owing to strong jet contribution to the observed UV/optical variability. 

We further explore the connection between radio emission and UV/optical variability in quasars from a different perspective. 
In Fig. \ref{Fig3_5_1} we plot the $g$-band $\sigma_{\mathrm{rms}}$ of the radio-detected quasars in the SDSS Stripe 82. 
We similarly build the control sample for them with matched redshift, $g$-band apparent magnitude, and $M_{\mathrm{bh}}$. Since we would like to measure $\sigma_{\mathrm{rms}}$ for the control sample as well, the control sample is selected from the SDSS Stripe 82 only. Due to the small sample size, for each radio-detected quasar we select only one control quasar. We remove a small fraction of the radio-detected quasars for which we can not find a control quasar in Stripe 82. 

In the upper panels of Fig. \ref{Fig3_5_1}, the clear correlation between the radio loudness and $g$-band $\sigma_{\mathrm{rms}}$ for the radio-detected quasars could be due to increasing jet contribution to the observed variability with increasing $R$. This may again explain the lack of excess ESQs at much larger $R$ (Fig. \ref{Fig3_3_1}), as ESQs could only be picked out of jetted quasars with low to intermediate $R$ for which jet contribution to the observed variability amplitude is weak. 

As shown in the lower panels of Fig. \ref{Fig3_5_1}, the radio-detected quasars have on average similar $g$-band $\sigma_{\mathrm{rms}}$ compared with the control sample. For the FIRST- and VLASS-detected quasars and CQs, the p-values of the KS test are 2.2\% and 13\%, respectively.
However, we see a clear excess of the radio-detected quasars at $\sigma_{\mathrm{rms}}$ $<$ 0.05 compared with the control sample (i.e., 57 ESQs vs 24 CQs for the FIRST detection, with a p-value of 0.0001, and 51 ESQs vs 24 CQs for the VLASS detection, with a p-value of 0.0009).
This is consistent with the pattern we have shown above, and suggests a link between jet production and ESQs.

To avoid strong jet contribution to the observed $g$-band $\sigma_{\mathrm{rms}}$, in Fig. \ref{fig-simulation} we plot the FIRST-detected quasars similar to the lower left panel of Fig. \ref{Fig3_5_1}
but excluding sources with $R > 60$. The statistical difference between the FIRST-detected quasars and the corresponding control sample is now evident (with a KS test p-value of 0.004), and the excess of the FIRST-detected sources with $g$-band $\sigma_{\mathrm{rms}} < 0.05$ remains significant (i.e., 46 ESQs vs 18 CQs, with a p-value of 0.0002).

Before discussing the physical implication of the radio excess in ESQs, there are two notable questions to be addressed: 1) why a large portion of ESQs are radio non-detected (Fig.~\ref{Fig3_1_1}), and 2) why some normal quasars could also have very small $g$-band $\sigma_{\mathrm{rms}}$ (the control sample in Fig.~\ref{Fig3_5_1})?

For the first question, we median-stack the radio images of radio non-detected ESQs and CQs \citep[see][for the stacking approach]{Liao2022}, 
and find higher median radio flux densities for ESQs than the control samples. Based on the FIRST (VLASS) images, the median radio flux density for ESQs is 92$^{+24}_{-18}$ $\mathrm{\mu}$Jy (76$^{+19}_{-20}$ $\mathrm{\mu}$Jy) and larger than 59$^{+5}_{-1}$ $\mathrm{\mu}$Jy (50$^{+5}_{-4}$ $\mathrm{\mu}$Jy) for the control sample. Though statistically marginal due to the small sample size of ESQs, this suggests that some of the radio non-detected ESQs also exhibit excess of radio emission. Much deeper radio images are desired to detect those radio fainter ESQs,
and address whether all true ESQs have jets.

For the second question, one factor we need to consider is the stochastic nature of the variation, that a normal variable quasar may exhibit large variation amplitude at one time, but much weaker variation at a different time. The sparse sampling and limited length of light curves could also play a role in magnifying such observational effect. We build mock light curves for the normal variable quasars assuming the damped random walk (DRW, \citealt{Kelly2009}) and non-DRW processes (see Appendix \ref{appendixE}). 
The simulations confirm that the normal variable quasars could temporarily appear as ``ESQs" due to the stochastic nature of the variation and/or the limited length of light curve, thus contaminate the selection of ESQs (see Fig. \ref{fig-simulation}).
The results of our simulations could also explain the drop of the radio-detected fraction of ESQs with increasing the threshold of $\sigma_{\mathrm{rms}}$ (see Fig. \ref{Fig3_1_1}), as the contamination from the normal variable quasars to ESQs is stronger at larger $\sigma_{\mathrm{rms}}$. 
Our simulations also show that longer light curves are essential to better distinguish real ESQs from the normal variable quasars (see Fig. \ref{fig-simulation} and Appendix \ref{appendixE}).

Now, it is intriguing to investigate how the prominent radio excess in ESQs reported here could be fitted in the most popular scenarios for the observed broad range of jet production efficiency in quasars and AGNs, i.e., the spin paradigm \citep[e.g.][]{Wilson1995,Sikora2007}.
It is commonly believed that the strong magnetic field is a key ingredient in the jet launching \citep[e.g.][]{BZ1977, BP1982, Livio1999,2019ARA&A..57..467B}, and the UV/optical variations in quasars could be driven by the disc turbulence induced by the magneto-rotational instability \citep[e.g.][]{Kelly2009}.
As proposed by \cite{Cai2019} presenting a tentative evidence for the more stable inner accretion disc in the RL quasars from the optical color variability study,
could the inner discs in the RL quasars have been stabilized by the strong magnetic field \citep[e.g.][]{Begelman2007,Zheng2011,Li2014,Sadowski2016}? 

Moreover, apart from the prominent radio excess in ESQs, another puzzle is why ESQs are so rare. 
Only $\sim$ 1.5\% of the quasars in the SDSS Stripe 82 (i.e., 135 out of 9146, based on the $\sigma_{\mathrm{rms}}$ threshold of 0.05 mag in all five SDSS bands) have been selected as ESQs, and only $\sim$ 7\% of the radio-detected quasars are ESQs (i.e., 34 out of 517 FIRST-detected ones, and 34 out of 499 VLASS-detected ones).
Assuming the magnetic field is a key factor on understanding the radio excess in ESQs as well as the rarity, we discuss below possible interpretations that could be further explored from both observational and theoretical aspects.

The rarity of ESQs may suggest that suppressing AGN variability by strong magnetic field in radio quasars could only occur in a minor fraction of sources
(see Figs. \ref{Fig3_5_1} \& \ref{fig-simulation}). {One possibility is that the suppressing is only significant with sufficiently strong magnetic field. This seems nicely in line with the theoretical analyses that the magneto-rotational instability (MRI) in the accretion disc could be stabilized beyond a critical magnetic field. Both \cite{Pessah2005} and \cite{Das2018} showed that a toroidal field with $\beta = 8 \pi P_{\rm gas}/ B^2 < 1$ (where $P_{\rm gas}$ is the gas pressure and $B$ the magnetic field) could suppress the MRI, while \cite{Bai2013} found a critical $\beta$ $<$ 100 for a poloidal field to suppress the MRI.
Therefore, we speculate that ESQs could be a minor population with magnetic field above a critical value (or $\beta$ below a critical value) in the inner disc. 
Note that the jet core-shift has been widely used to measure the magnetic field in jet at pc scale \citep[e.g.][]{Lobanov1998,Zamaninasab2014}. While a quantitative ratio between magnetic field in the inner disc and that in the pc-scale jet is unclear, it would be helpful to investigate with future core-shift observations whether ESQs have stronger pc-scale magnetic field compared with the control sample.

In the spin diagram, the jet power, $P_{\rm jet}$, depends on both the black hole spin and the poloidal magnetic field ($P_{\rm jet} \sim \Omega_H^2B_p^2$, where $\Omega_H$ is the angular velocity at the black hole horizon and $B_p$ the poloidal magnetic field; e.g.,  \citealt{Livio1999}). 
Under this scheme, the proposed beyond-critical magnetic field
could naturally explain the excess of the radio emission in ESQs. 
Meanwhile, to explain their intermediate radio loudness ($R \sim 10 - 60$), ESQs should have relatively small black hole spins resulting in small $\Omega_H$ thus moderate $P_{\rm jet}$. These ESQs would belong to the dominant radio quiet population if without the beyond-critical magnetic field. 
Contrarily, some ESQs with both high spin and beyond-critical poloidal magnetic field could be very radio loud. However they could be even rarer compared with the radio intermediate ESQs, and their optical/UV variability could be significantly elevated due to the jet contribution, thus missed by our selection of ESQs.
On the other hand, if there are ESQs with strong toroidal field but too weak poloidal field, jet launching would not be expected \citep{Beckwith2008}, which may also explain some of the radio non-detected ESQs.  
}

Overall, our results and the above plausible interpretations make ESQs an essential and unique population under potentially extreme condition, which strongly necessitates extensive theoretical and observational follow-up studies. 
The quantitative relation between strong magnetic field suppressing and the observed AGN variability is to be established theoretically.  
Much deeper radio images could draw a panorama of the radio loudness of ESQs. 
Multi-band radio SEDs and high-resolution radio images could reveal the physical nature of the jets in ESQs, to verify the existence of strong magnetic field.
In additional to follow-up radio observations, our work could also makes ESQs unique targets for the time domain astronomy, because the longer duration/higher cadence sampling of time domain observations, the better that stableness of a quasar could be  constrained. 
Studying the relation between ESQs and other multi-band observed parameters of quasars may also yield new hints (see Appendix~\ref{appendixG}, Figs.~\ref{fig_appendixG} and \ref{fig_appendixG2} for instance).
Extending the study (of the connection between disc stableness and jet) to the low mass regime, i.e, stellar mass BHs, is also alluring.

\section*{Acknowledgement}
We thank the anonymous referee for his/her helpful comments.
The work is supported by National Key R\&D Program of China No. 2023YFA1607903, National Natural Science Foundation of China (grant nos. 12033006, 12373016, 12192221, and 11890693). 
ZYC is supported by the science research grants from the China Manned Space Project under grant no. CMS-CSST-2021-A06 and the Cyrus Chun Ying Tang Foundations.
FY acknowledges the support from NSFC grant 12133008, 12192220, and 12192223. AAZ has been supported by the Polish National Science Center under the grant 2019/35/B/ST9/03944.

\appendix
\section{A: Weighted excess variance}\label{appendixA}

\begin{figure}
  \centering
  \includegraphics[width=\linewidth]{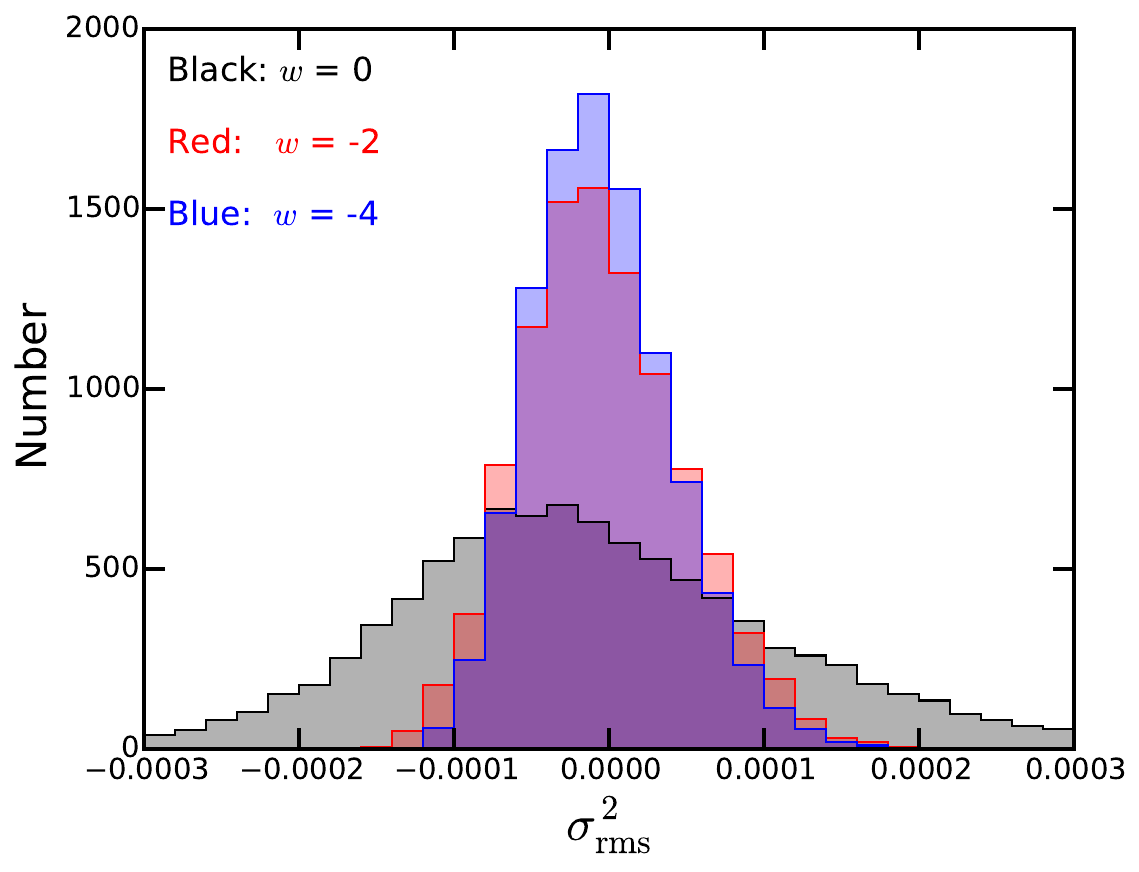}
  \caption{Distribution of the simulated $\sigma_{\mathrm{rms}}^2$ (zero intrinsic variance is assumed) using different weighted factors.  
}
  \label{Fig2_1_1}
\end{figure}

We calculate the revised $\sigma_{\mathrm{rms}}^2$ for a quasar light curve with $N$ epochs as 
{
 \begin{align}
  \sigma_{\mathrm{rms}}^2=\frac{1}{\sum \sigma^w_i}\sum \sigma^w_i[\frac{N}{N-1} (X_i-\bar{X})^2 - \sigma_i^2],
   \label{e2_1_1}
 \end{align}
}
where $X_i$ the observed magnitude at $i$ epoch, $\sigma_i$ is the photometric uncertainty at $i$ epoch, and
$\bar{X}$ is the weighted average magnitude in which the weight is $\sigma^2_i$. We consider three different weighted factors $\sigma_{i}^w$, with $w =$ 0, -2, and -4, respectively. When $w$ = 0, the equation returns to the canonical form \citep{2003MNRAS.345.1271V}. The smaller  (more negative) $w$ is, the smaller the influence of the observed epochs with large photometric uncertainties to the variability amplitude.

Utilizing the SDSS $g$ band as an example, we perform Monte Carlo simulations to select the optimal weighted factor for the selection of ESQs with very weak intrinsic variability. 
Suppose a quasar with zero intrinsic variability and $g$ = 19.5 has been observed $n$ times, we simulate its magnitude at each epoch as:
 \begin{align}
  g_i = g + Gau(0, \sigma_i) 
   \label{e2_1_2}
 \end{align}
where $\sigma_i$ refers to the photometric uncertainty. We adopt $n$ = 60 which is the average observation number for each source in the SDSS Stripe 82 \citep{Macleod2012}. 
To mimic the true observational effects, $\sigma_i$ is randomly selected from the real data (out of all photometric measurements of the SDSS Stripe 82 quasars with observed $g = 19.5 \pm 0.1$ at any epoch).
Distribution of the simulated $\sigma_{\mathrm{rms}}^2$ is displayed in Fig. \ref{Fig2_1_1}. 
Clearly, the canonical form of $\sigma_{\mathrm{rms}}^2$ ($w$ = 0) yields the largest scatter, while $w$ = -4 behaves slightly better than $w$ = -2. Simulations assuming various $g$ band magnitudes yield similar results. 
Thus in this work, we adopt $w$ = -4 to calculate $\sigma_{\mathrm{\mathrm{rms}}}^2$.

We also perform similar Monte Carlo simulations to derive the statistical uncertainty of a measured $\sigma_{\mathrm{rms}}^2$, i.e., \emph{error}($\sigma_{\mathrm{rms}}^2$), assuming a quasar has zero intrinsic variability and estimating the expected standard deviation of $\sigma_{\mathrm{rms}}^2$ (such as the scatter of the mock $\sigma_{\mathrm{rms}}^2$ plotted in Fig. \ref{Fig2_1_1}) simply due to photometric errors. Note this approach is only valid for sources with extremely weak variability, but sufficient for this study.

\section{B: build the control samples}\label{appendixB}

For each ESQ, we randomly select 10 control quasars out of the SDSS DR 16 quasar catalog  \citep{Lyke2020} with $\Delta z/z$ $\leq$ 10\%, $\Delta g$ $\leq$ 0.2 mag, and $\Delta \log M_{\mathrm{bh}}$ $\leq$ 0.4 dex. 
The reason we select 10 (instead of one) CQs for each ESQ is to reduce Poisson noise of the control sample. 
Note while the VLASS fully covers the SDSS survey, the FIRST does not. We only select CQs within the FIRST footprint, and exclude one ESQ which is out of the FIRST footprint from Fig. \ref{Fig3_1_1} when presenting the FIRST detection fraction.  
We further drop one more ESQ which has less than 5 CQs available. For the two other ESQs with at least 5 but less than 10 CQs, their available CQs are repeatedly and randomly selected to make up 10 CQs.  
 
We further note the FIRST sensitivity is slightly deeper in a small region along the equatorial strip (RA = 21.3 to 3.3 hr, Dec = -1 to 1 deg, with a typical detection threshold of 0.75 mJy, instead of 1.0 mJy for the rest dominant FIRST survey area). The deeper strip is right within the SDSS Stripe 82 field. While comparing the FIRST detection fraction of ESQs with the control samples, we need to correct the effect of the non-uniform depth of the FIRST survey. To do so, for any ESQ within the deeper strip, we select its CQs from the DR16Q but with $g$-band flux density brighter by a factor of 1/0.75 to compensate the effect (slightly brighter CQs are needed to be detected in slightly shallower FIRST image). We also request the control quasar to have BH mass larger by a factor of 1/0.75 to ensure the match of Eddington ratio, as the RL fraction could be dependent on the Eddington ratio. After this correction, the FIRST detection fraction of the CQs only slightly increases (by around 1\%) and has a negligible effect to the results presented in this study. 
Note the results we present in Fig. \ref{Fig3_1_1} are already after this correction. Furthermore, the FIRST detection fraction of our ESQs in and out the deeper FIRST strip exhibits no significant difference, also indicating the effect of the non-uniform FIRST depth is negligible.    

\section {C: Cross-match the SDSS DR16Q with the radio catalogs}\label{appendixD}

\begin{figure}
  \centering
  \includegraphics[width=8cm]{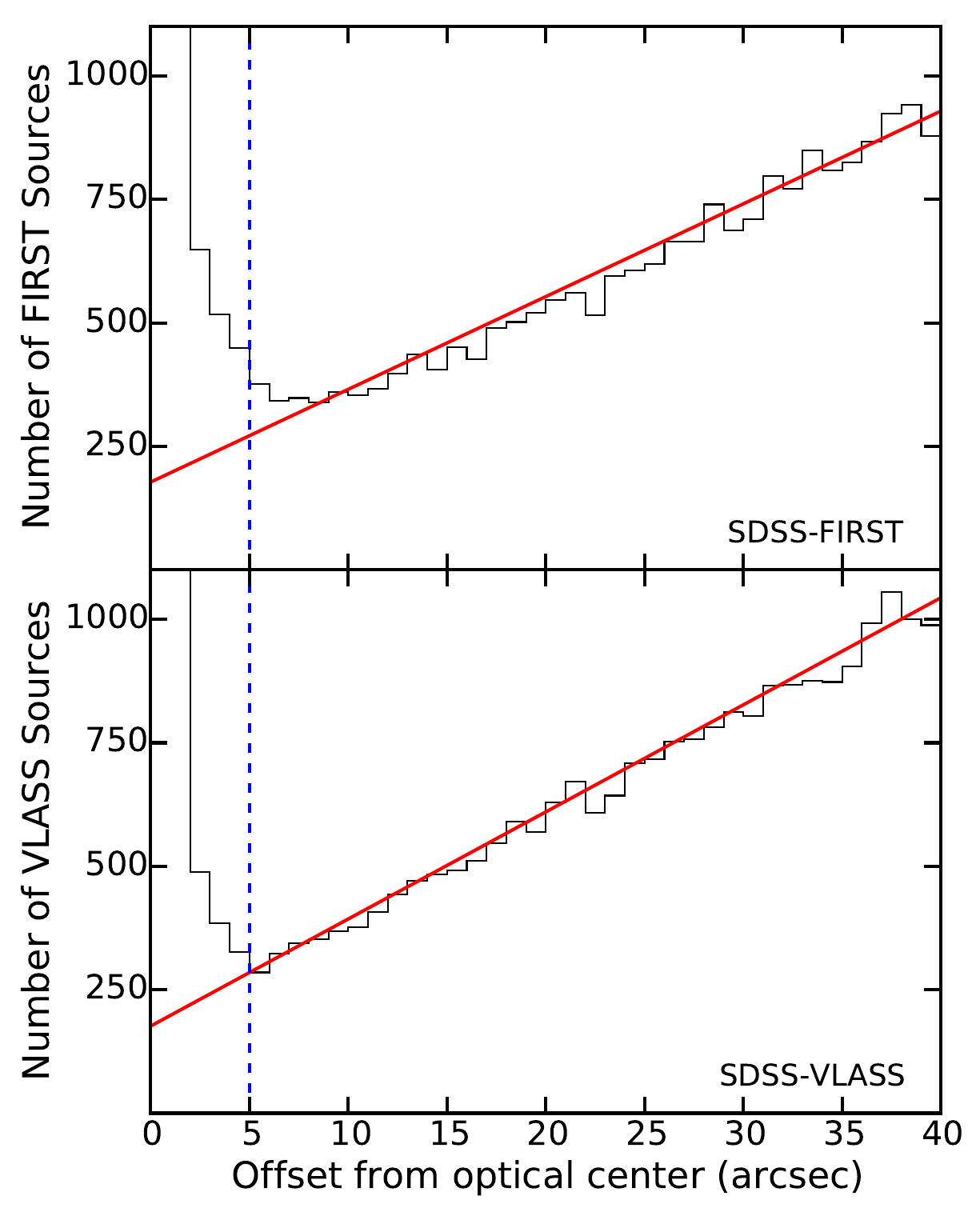}
  \caption{ The observed offset distribution of cross-matching the SDSS DR16Q catalog with the FIRST (upper panel) and VLASS catalogs (lower panel).
  We adopt 5\arcsec\ as the cross-matching threshold (blue dashed lines). The red lines show the expected number of chance alignments with background sources as a function of offset between SDSS and radio sources.}
  \label{fig_radius}
\end{figure}

\begin{figure}
  \centering
  \includegraphics[width=8cm] {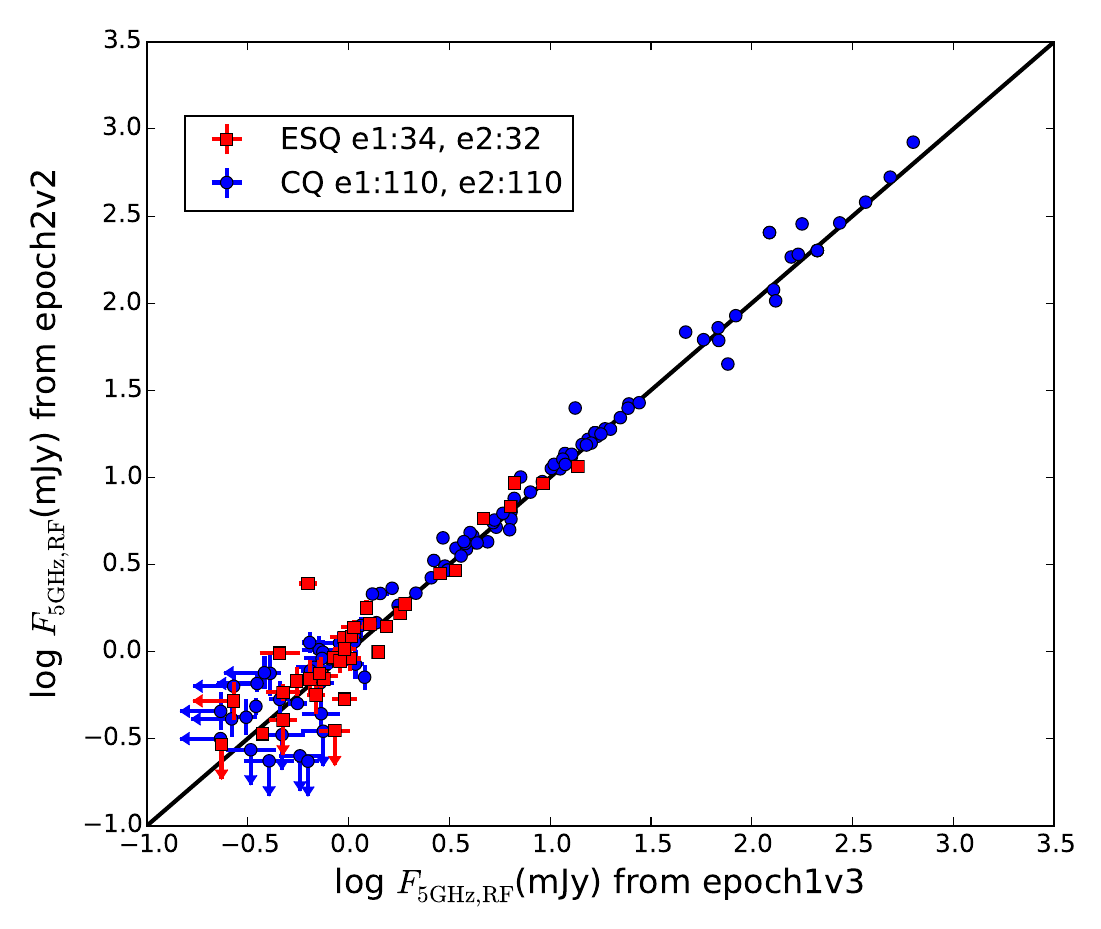}
  \caption{ Radio flux densities from the VLASS epoch 1 (version 3) vs epoch 2 (version 2).  Both of them have been converted to the rest-frame 5 GHz assuming a spectral index of $\alpha$ = 0.5. Numbers of the VLASS-detected sources are marked. For a few sources detected in only one epoch, the upper limits are given for the other epoch.  
  }
  \label{fig_appendixD3}
\end{figure}

\begin{figure}
  \centering
  \includegraphics[width=\linewidth]{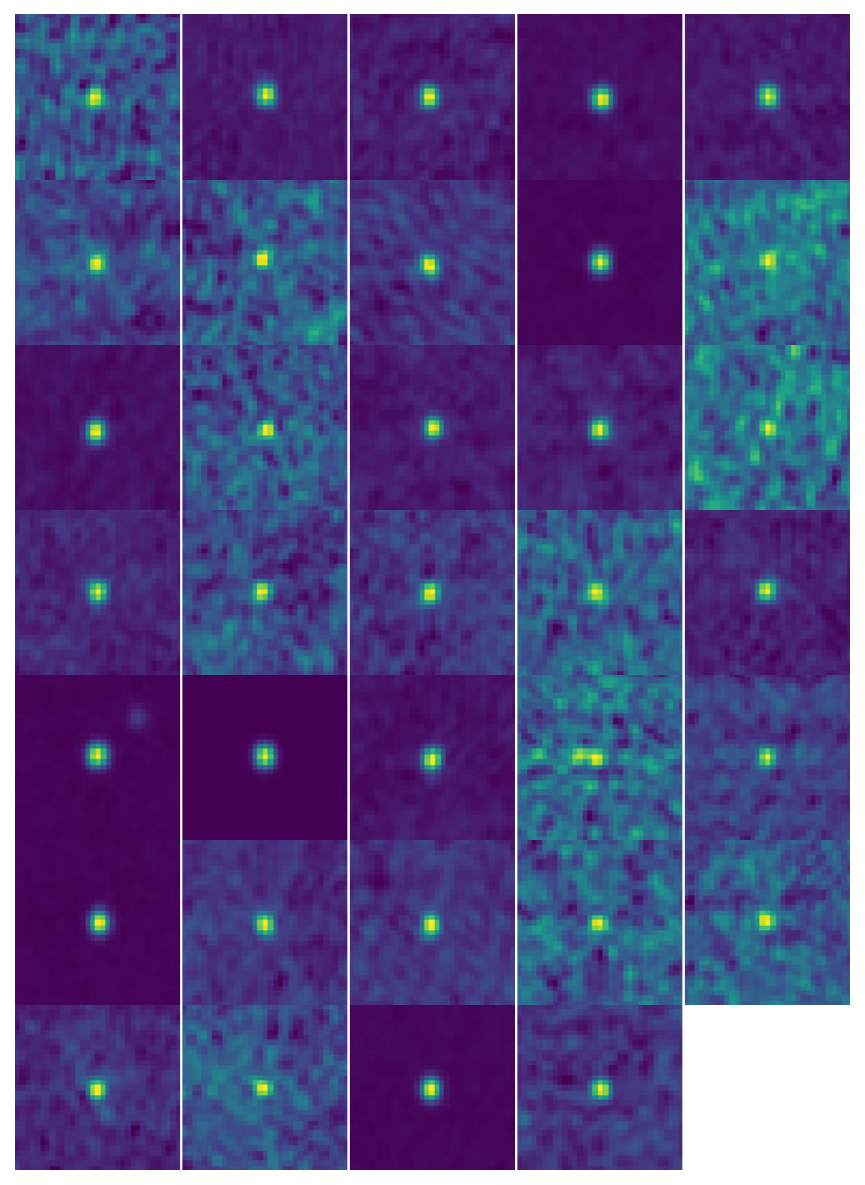}
  \caption{ The FIRST cutout images (60\arcsec$\times$60\arcsec) of the 34 FIRST-detected ESQs. In one of these cutouts images, i.e., the fourth column from the left and the third row from the bottom, there seems to be two radio counterparts. However only one radio source (in the center of the cutout, $\simeq 0.1''$ offset from SDSS position) is detected in the FIRST catalog, and the other point-like source to the slightly left is likely due to jet structure.  }
  \label{fig_FIRSTimageESQ}
\end{figure}

\begin{figure}
  \centering
  \includegraphics[width=\linewidth]{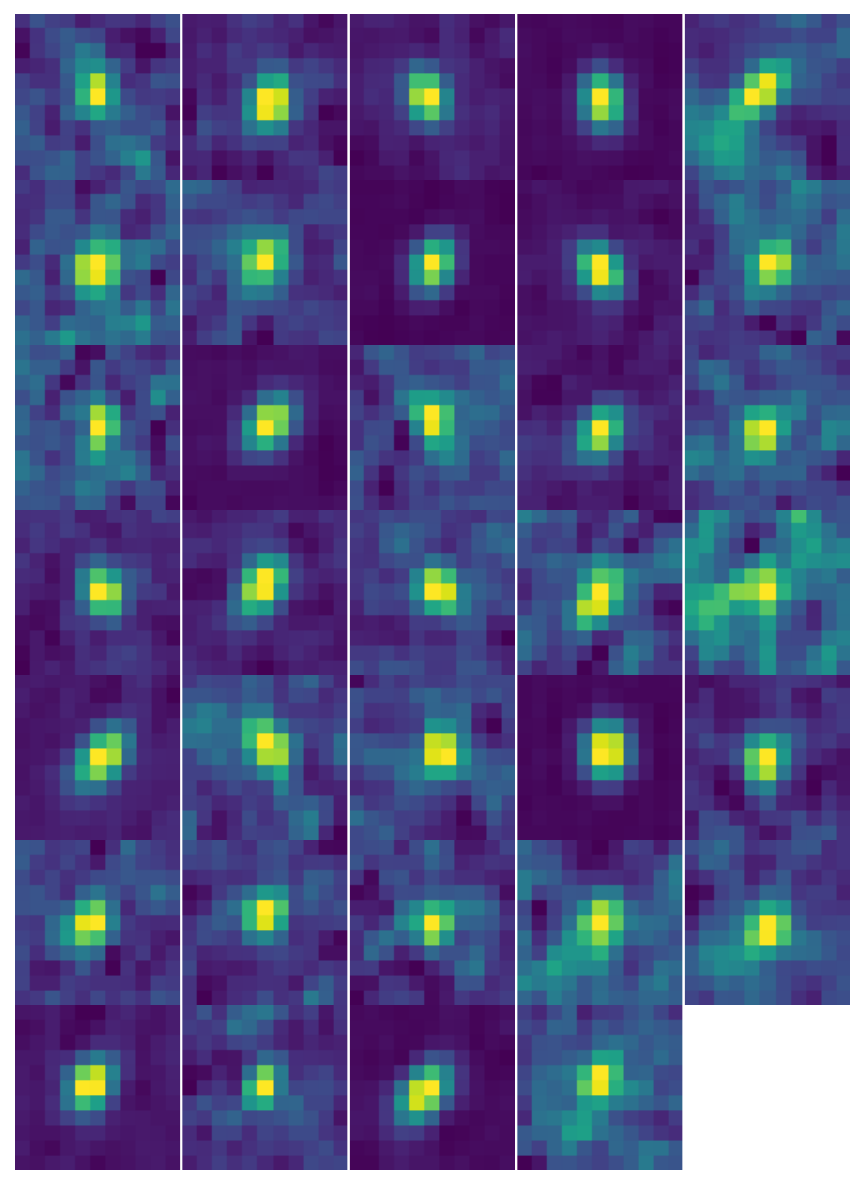}
  \caption{ The VLASS epoch 1 cutout images (11\arcsec$\times$11\arcsec) of the 34 VLASS-detected ESQs.}
  \label{fig_VLASSimageESQ}
\end{figure}

\begin{figure*}
  \centering
  \includegraphics[width=\linewidth]{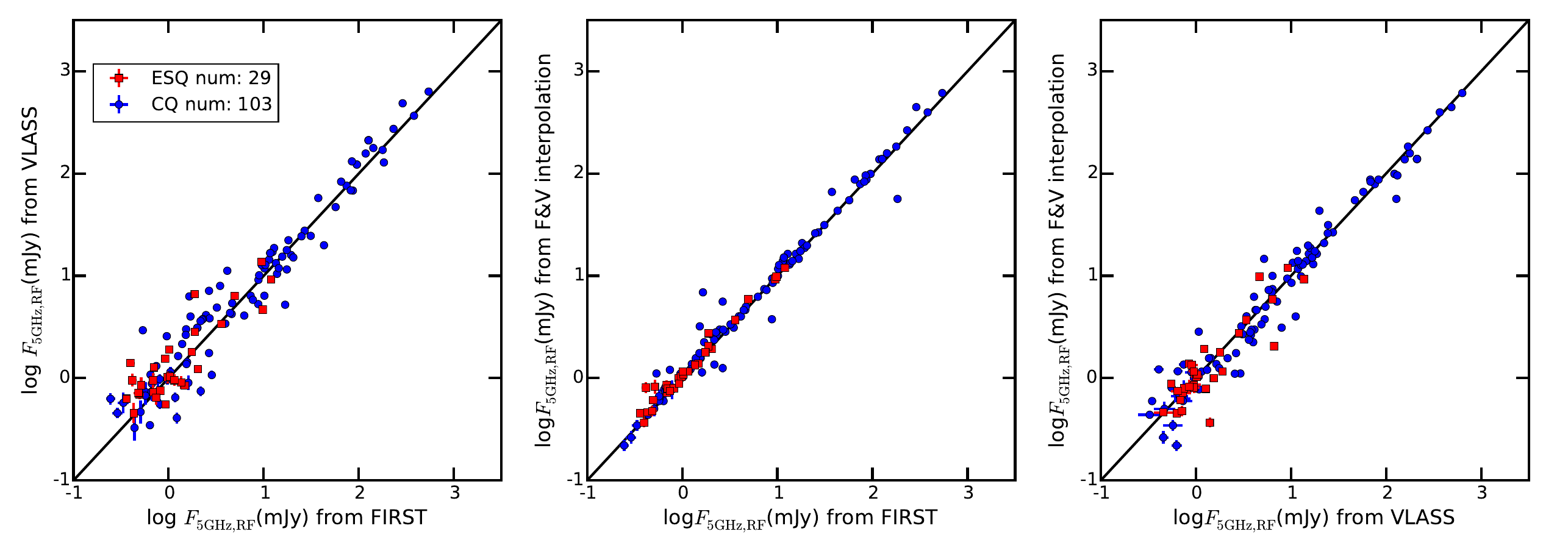}
  \caption{ Left: Rest-frame 5GHz flux density of ESQs and CQs (detected by both FIRST and VLASS) derived from the observed frame 3 GHz VLASS  flux density ($y$-axis) versus that derived from the 1.4 GHz FIRST flux density ($x$-axis), both assuming a spectral index of $\alpha$ = 0.5. Middle and Right: Rest-frame 5GHz flux density of ESQs and CQs derived through interpolating FIRST and VLASS flux densities ($y$-axis) versus that from FIRST and VLASS ($x$-axis of middle right panel, respectively, both assuming spectral index of $\alpha$ = 0.5).}
  \label{fig_appendixD2}
\end{figure*}

We cross-match the SDSS DR16Q with both the FIRST and VLASS catalogs, using a matching radius of 5$\arcsec$. Following \cite{2020ApJ...888...36R}, in Fig. \ref{fig_radius} we plot the offset distribution from cross-matching the SDSS DR16Q catalog with the radio catalogs. As indicated by the red lines, the expected fraction of chance alignments with background sources for the radio-detected quasars below 5\arcsec\ offset is 4.6\% (1082 out of 23458 FIRST detected ones) and 4.3\% (1104 out of 25782 VLASS detected ones), respectively. Our Fig.~\ref{fig_radius} is similar to Fig. 2 of \cite{2020ApJ...888...36R}.

For the VLASS, we adopt the epoch 1 Quick Look catalog \citep{Gordon2021}. Utilizing the epoch 2 catalog does not alter our results. Meanwhile, comparing the epoch 1 and epoch 2 VLASS flux densities of our ESQs or the control sample reveals no systematic variation trend (see Fig.~\ref{fig_appendixD3}). 

In Fig.~\ref{fig_FIRSTimageESQ} (and Fig.~\ref{fig_VLASSimageESQ}) we illustrate the FIRST (VLASS) images of the ESQs with FIRST (VLASS) detection.  
We also compare the rest frame 5 GHz flux densities derived from the FIRST or VLASS assuming a spectral index of 0.5, and from interpolating the FIRST and VLASS flux densities (Fig.~\ref{fig_appendixD2}). We find that these approaches yield generally consistent rest frame flux densities. Fig.~\ref{fig_appendixD2} also indicates that the observed frame 1.4 -- 3 GHz spectral slopes of ESQs and their control sample exhibit no systematic difference. 

\section{D: the color of ESQs vs the control samples}\label{appendixC}

\begin{figure}
  \centering
  \includegraphics[width=\linewidth]{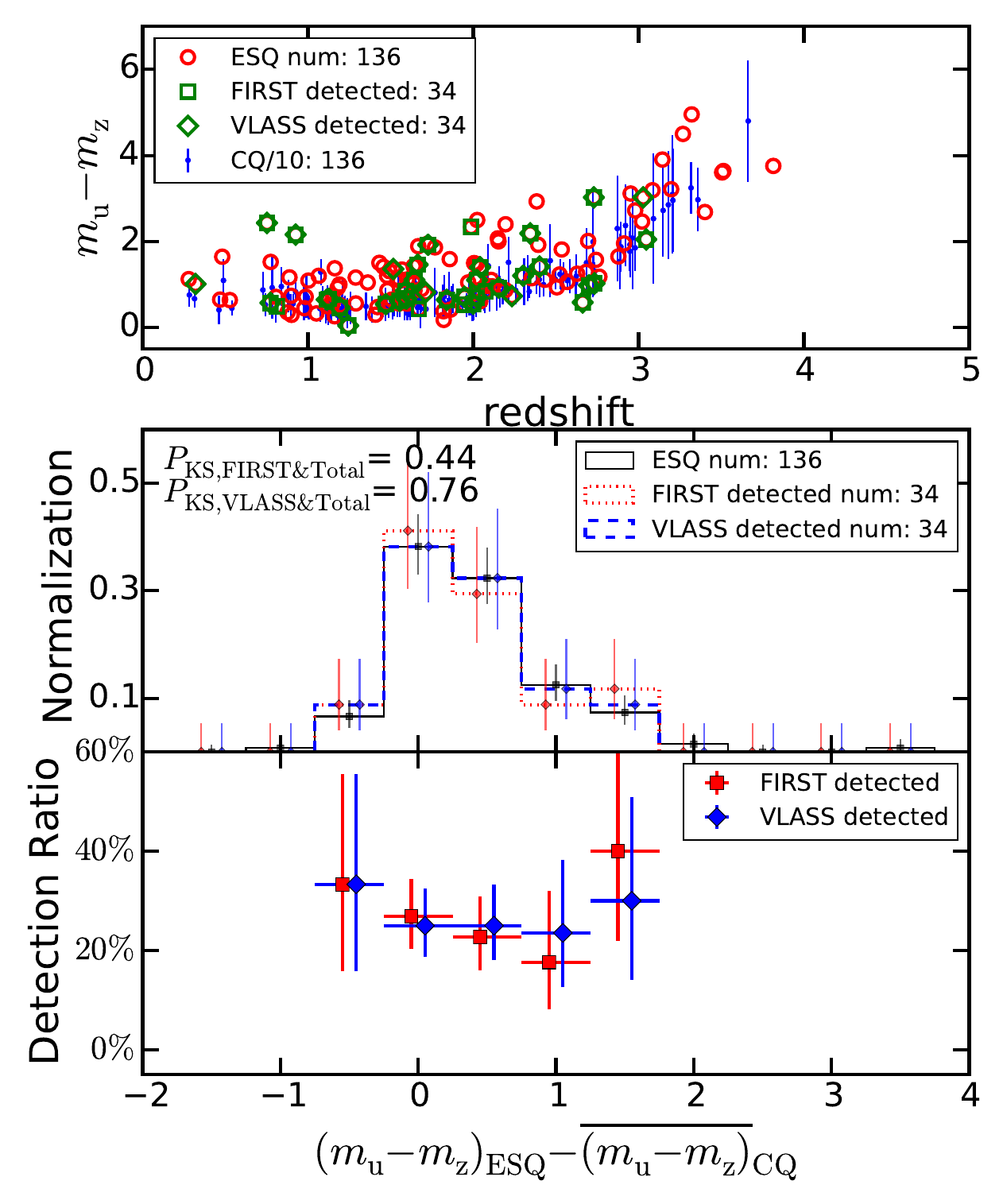}
  \caption{Upper: the $u$-$z$  color of our ESQs (selected with a threshold of $\sigma_{\mathrm{rms}} \leq 0.05$) versus that of the control sample.  Red circles are ESQs while green squares and diamonds are the FIRST and VLASS detected ESQs, respectively. The statistical errors to the colors due to photometric uncertainties are generally too small to be displayed. 
  For each ESQ we have 10 CQs, and we also plot the mean color (and standard deviation) of the 10 CQs for comparison.
  Middle: the histogram distribution of the color difference between ESQs and the control quasars.  Error bars are 1$\sigma$ confidence limits for Poisson statistics. The $P_{\mathrm{KS}}$ shown in the upper-left corner are the p-values of the KS test between ESQs and FIRST detected ESQs, and between ESQs and VLASS detected ESQs, respectively.
  Lower: the radio detection fraction of ESQs versus the color difference.  Error bars are 1$\sigma$ confidence limits based on a combination of Poisson and binomial statistics.
}
  \label{fig_color}
\end{figure}

In Fig. \ref{fig_color} we plot the $u$-$z$ color of ESQs compared with the control samples. While a considerable fraction of ESQs do exhibit redder $u$-$z$ color, the radio detection fraction in ESQs appears independent to the color. 

\section{ E: contamination of normal quasars to ESQs}\label{appendixE}

We perform simulations to assess the contamination to ESQs by the normal variable quasars appearing as ``ESQs'' as a result of effects of the stochastic nature of variation (i.e., the DRW and non-DRW processes with distinct damping timescales), the limited length of light curve (i.e., 3 years, 10 years, and 30 years), and the sparse sampling. In order to compare with the $g$-band $\sigma_{\rm rms}$ distributions of the FIRST detected quasars with $R < 60$ and the corresponding control sample, we implement their real $g$-band samplings in our simulations. Both of them have 303 quasars.

For the DRW process, light curves are simulated using the procedure of \citet[][]{Kelly2009} with three typical damping timescales ($\tau_{\rm DRW} = $ 100 days, 300 days, and 1000 days) and a long-term variation amplitude ($\sigma_{\rm l} = 0.1$ mag) for the normal variable quasars.
For the non-DRW process, light curves are simulated in terms of the algorithm of \citet{Timmer1995A&A...300..707T} assuming a broken power-law power spectral density (PSD). The non-DRW PSD breaks at a frequency of $(2 \pi \tau_{\rm non-DRW})^{-1}$ with $\tau_{\rm non-DRW} = 1000$ days, and has a low-frequency slope $\alpha_{\rm l} = -1.3$ suggested by \citet{Guo2017ApJ...847..132G} and a high-frequency slope $\alpha_{\rm h} = -2$. All light curves are generated in steps of 0.1 day. The length of the DRW light curves are set to 30 years, while $\simeq 90$ years for the non-DRW light curves, whose long-term variation amplitudes are $\sigma_{\rm l} = 0.1$ mag as well. A longer length for the non-DRW process is to fully take the diverse power at low frequencies into account; and a length longer than $\simeq 90$ years for the non-DRW process does not alter the results implied by $\simeq 90$-year light curves.

For each quasar, we generate 1000 light curves, which are coupled with its real $g$-band sampling. Specifically, each simulated light curve is linearly interpolated at the observed epochs of a quasar, and the interpolated magnitudes are fluctuated by random Gaussian deviates scaled to the observed photometric uncertainties of the quasar. The length of the observed light curves for quasars is 10-year long. The 10-year real observation is repeated 3 times to mimic a 30-year observation, while the last 3-year of the 10-year real observation mocks a 3-year observation. 

In Fig.~\ref{fig-simulation} we compare the $g$-band $\sigma_{\rm rms}$ distributions for both the RD quasars with $R < 60$ and the corresponding CQ with those simulated for a single normal variable quasar, considering different damping timescales of the DRW process, the effect of the non-DRW process, and the sampling lengths. 
For each of these effects, the simulated ``absolute'' distributions of the $g$-band $\sigma_{\rm rms}$ also depend on the adopted $\sigma_{\rm l}$.
However, adopting different input $\sigma_{\rm l}$ would just horizontally shift the output distribution of the $g$-band $\sigma_{\rm rms}$ without changing its shape.
The similarity of the ``relative'' distributions for various input $\sigma_{\rm l}$ is confirmed by simulations using a series of $\sigma_{\rm l}$, ranging from 0.01 to 0.4, well blanketing the observed values. Therefore, we adopt the typical $\sigma_{\rm l} = 0.1$ for reference and normalized (and shifted horizontally) the resultant distributions of the $g$-band $\sigma_{\rm rms}$ to around the peak of the observed CQ distribution (Fig.~\ref{fig-simulation}).

As expected, longer damping timescale, the non-DRW process rather than the DRW one, and shorter sampling length can all increase the probability of a normal variable quasar temporarily appearing as an ``ESQ". As shown in Fig.~\ref{fig-simulation}, there are 46 RD quasars and 18 CQs with $g$-band $\sigma_{\rm rms} < 0.05$. Assuming the most broad distribution simulated by the non-DRW process with $\tau_{\rm non-DRW} = 1000$ days and $\alpha_{\rm l} = -1.3$, $\sim 13\%$ of the RD quasars and $\sim 35\%$ of the CQs are likely normal variable quasars temporarily appearing as ``ESQs".
Note that the fractions of contamination by the normal variable quasars are hard to assess as the exact variation model (i.e., DRW vs non-DRW), the model parameters (e.g., the damping timescale and the low frequency PSD slope) and the variance from source to source (currently omitted in the simulations) are yet poorly constrained \citep[e.g.][]{Guo2017}. Even longer damping timescale (which is very likely as the measured timescale could have been under-estimated due to the limited length of light curves, e.g., \citealt{HuXF2023}) would yield higher contamination by the normal variable quasars (see the left panel of Fig. \ref{fig-simulation}). Meanwhile, longer light curves appear essential to better distinguish real ESQs from the normal variable quasars (see the right panel of Fig. \ref{fig-simulation}).

\section{ F: Other factors}\label{appendixG}

\begin{figure}
  \centering
  \includegraphics[width=\linewidth]{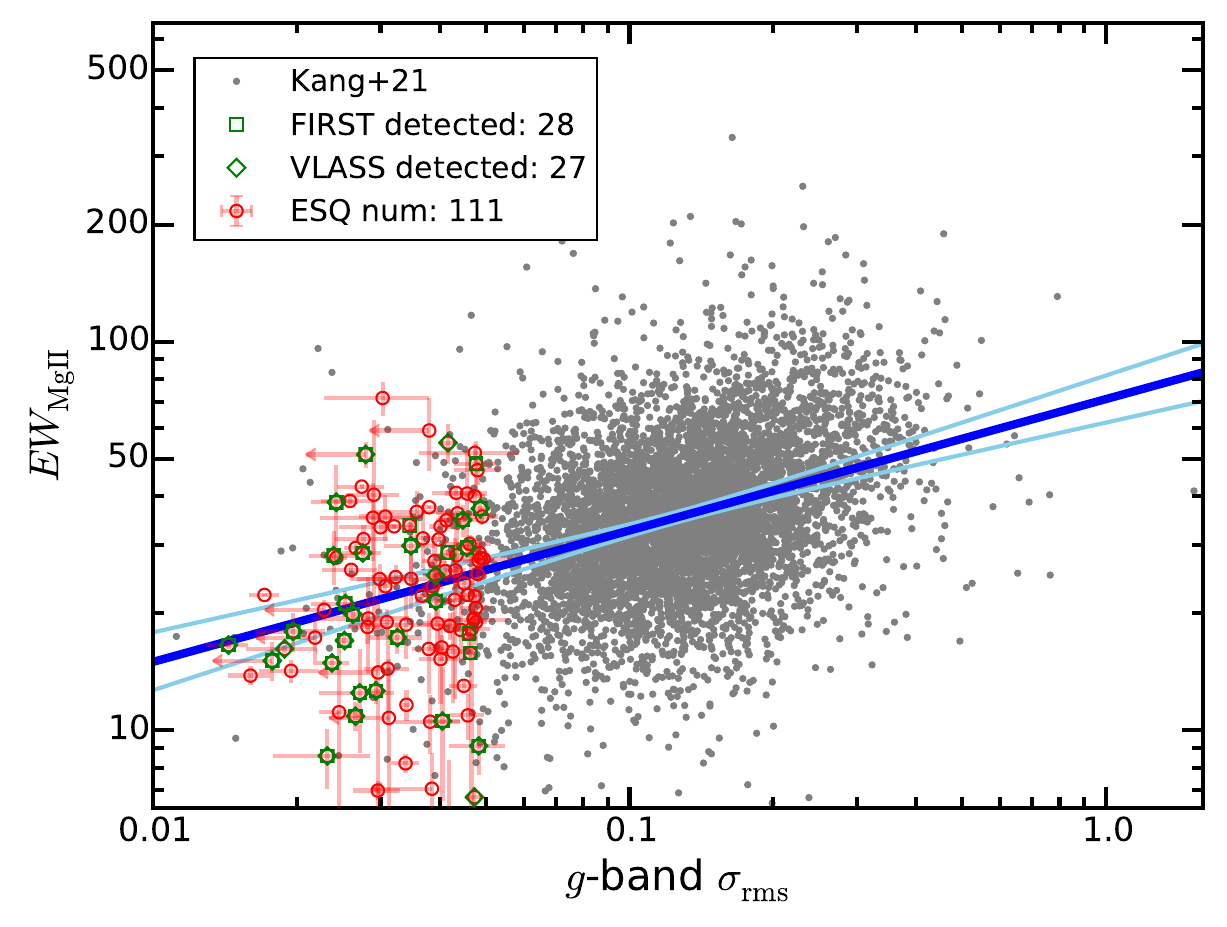}
  \caption{ Distribution between the equivalent width ($\rm EW$) of broad Mg II line and the $g$-band $\sigma_{\mathrm{rms}}$. Blue lines and grey points are results from \cite{2021ApJ...911..148K}. The red circles are our ESQs. The green squares and diamonds are the FIRST and VLASS detected ESQs, respectively.
}
  \label{fig_appendixG}
\end{figure}

\begin{figure}
  \centering
  \includegraphics[width=\linewidth]{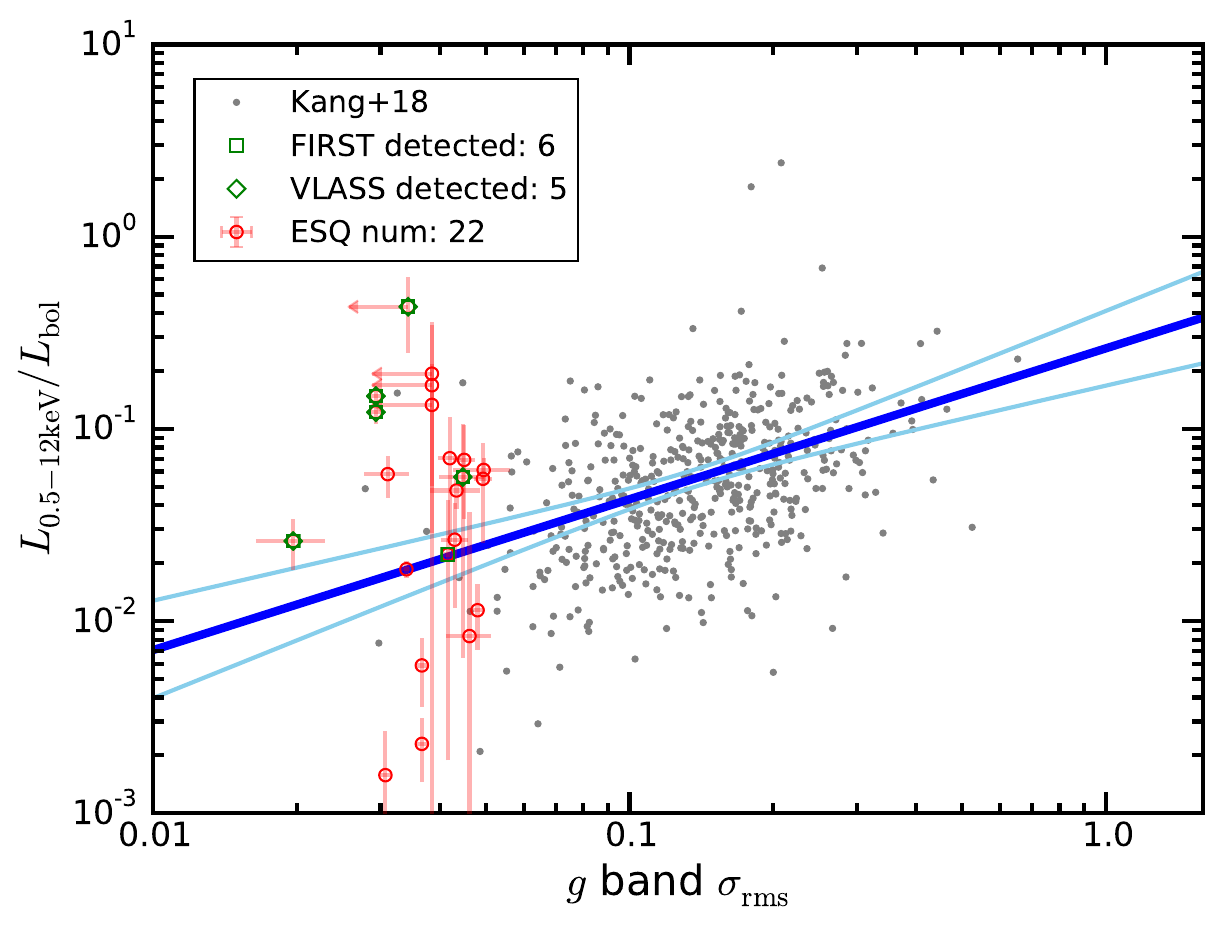}
  \caption{ Distribution between the X-ray loudness of 0.5 - 12 keV (from the 4XMM-DR13 catalog; \citealt{2020A&A...641A.136W}) and the $g$-band $\sigma_{\mathrm{rms}}$. The blue lines and grey points are results from \cite{2018ApJ...868...58K}. The red points are our ESQs. The green squares and diamonds are the FIRST and VLASS detected ESQs, respectively.
}
  \label{fig_appendixG2}
\end{figure}

In the manuscript we have controlled the effects of redshift, luminosity, BH mass, and Eddington ratio while comparing ESQs with the control sample. Here we present attempts to explore the effects of other potential factors. 

In \cite{2021ApJ...911..148K}, we have found that the UV/optical variability amplitude of quasars positively correlates with the equivalent widths (EWs) of C IV, Mg II, and [O III]5007 emission lines. In Fig.~\ref{fig_appendixG}, we plot our ESQs which have EWs of broad Mg II line from the SDSS DR16Q catalog \citep{Wu2022}, compared with quasars from Fig.~4 of \cite{2021ApJ...911..148K}. Here the linear regression is obtained taking $g$-band $\sigma_{\mathrm{rms}}$ as the independent variable, used to illustrate the expected $EW_{\mathrm{MgII}}$ at given $g$-band $\sigma_{\mathrm{rms}}$. We see that the ESQs are consistent with the correlation that was fit to the \cite{2021ApJ...911..148K} sample, suggesting that compared with the normal variable quasars the ESQs are not distinct in their emission line strength.  

The UV/optical variability of quasars was also found be correlate with the X-ray loudness of quasars \citep{2018ApJ...868...58K}. We cross-match our ESQs with the XMM-DR13 catalog \citep{2020A&A...641A.136W}, and compare our ESQs with the normal variable quasars from \cite{2018ApJ...868...58K} in the X-ray loudness versus $g$-band $\sigma_{\mathrm{rms}}$ (Fig.~\ref{fig_appendixG2}). Similarly we find that the ESQs are consistent with the correlation that was fit to the \cite{2018ApJ...868...58K} sample. Meanwhile, the radio-detected ESQs tend to be X-ray louder (statistically yet insignificant due to the small number of ESQs with X-ray coverage).

\bibliography{ref}

\begin{thebibliography}{}
\providecommand\natexlab[1]{#1}
\providecommand\JournalTitle[1]{#1}

\bibitem[{{Ai} {et~al.}(2010){Ai}, {Yuan}, {Zhou}, {Wang}, {Dong}, {Wang}, \&
  {Lu}}]{2010ApJ...716L..31A}
{Ai}, Y.~L., {Yuan}, W., {Zhou}, H.~Y., {et~al.} 2010,
  \href{http://dx.doi.org/10.1088/2041-8205/716/1/L31}{\JournalTitle{\apjl},
  716, L31}

\bibitem[{{Bai} \& {Stone}(2013)}]{Bai2013}
{Bai}, X.-N., \& {Stone}, J.~M. 2013,
  \href{http://dx.doi.org/10.1088/0004-637X/769/1/76}{\JournalTitle{\apj}, 769,
  76}

\bibitem[{{Bauer} {et~al.}(2009){Bauer}, {Baltay}, {Coppi}, {Ellman}, {Jerke},
  {Rabinowitz}, \& {Scalzo}}]{Bauer2009}
{Bauer}, A., {Baltay}, C., {Coppi}, P., {et~al.} 2009,
  \href{http://dx.doi.org/10.1088/0004-637X/696/2/1241}{\JournalTitle{\apj},
  696, 1241}

\bibitem[{{Beckwith} {et~al.}(2008){Beckwith}, {Hawley}, \&
  {Krolik}}]{Beckwith2008}
{Beckwith}, K., {Hawley}, J.~F., \& {Krolik}, J.~H. 2008,
  \href{http://dx.doi.org/10.1086/533492}{\JournalTitle{\apj}, 678, 1180}

\bibitem[{{Begelman} \& {Pringle}(2007)}]{Begelman2007}
{Begelman}, M.~C., \& {Pringle}, J.~E. 2007,
  \href{http://dx.doi.org/10.1111/j.1365-2966.2006.11372.x}{\JournalTitle{\mnras},
  375, 1070}

\bibitem[{{Blandford} {et~al.}(2019){Blandford}, {Meier}, \&
  {Readhead}}]{2019ARA&A..57..467B}
{Blandford}, R., {Meier}, D., \& {Readhead}, A. 2019,
  \href{http://dx.doi.org/10.1146/annurev-astro-081817-051948}{\JournalTitle{\araa},
  57, 467}

\bibitem[{{Blandford} \& {Payne}(1982)}]{BP1982}
{Blandford}, R.~D., \& {Payne}, D.~G. 1982,
  \href{http://dx.doi.org/10.1093/mnras/199.4.883}{\JournalTitle{\mnras}, 199,
  883}

\bibitem[{{Blandford} \& {Znajek}(1977)}]{BZ1977}
{Blandford}, R.~D., \& {Znajek}, R.~L. 1977,
  \href{http://dx.doi.org/10.1093/mnras/179.3.433}{\JournalTitle{\mnras}, 179,
  433}

\bibitem[{{Cai} {et~al.}(2019){Cai}, {Sun}, {Wang}, {Zhu}, {Gu}, \&
  {Yuan}}]{Cai2019}
{Cai}, Z., {Sun}, Y., {Wang}, J., {et~al.} 2019,
  \href{http://dx.doi.org/10.1007/s11433-018-9330-4}{\JournalTitle{Science
  China Physics, Mechanics, and Astronomy}, 62, 69511}

\bibitem[{{Cohen} {et~al.}(1986){Cohen}, {Rudy}, {Puetter}, {Ake}, \&
  {Foltz}}]{Cohen1986}
{Cohen}, R.~D., {Rudy}, R.~J., {Puetter}, R.~C., {Ake}, T.~B., \& {Foltz},
  C.~B. 1986, \href{http://dx.doi.org/10.1086/164758}{\JournalTitle{\apj}, 311,
  135}

\bibitem[{{Das} {et~al.}(2018){Das}, {Begelman}, \& {Lesur}}]{Das2018}
{Das}, U., {Begelman}, M.~C., \& {Lesur}, G. 2018,
  \href{http://dx.doi.org/10.1093/mnras/stx2518}{\JournalTitle{\mnras}, 473,
  2791}

\bibitem[{{Dunlop} {et~al.}(2003){Dunlop}, {McLure}, {Kukula}, {Baum}, {O'Dea},
  \& {Hughes}}]{Dunlop2003}
{Dunlop}, J.~S., {McLure}, R.~J., {Kukula}, M.~J., {et~al.} 2003,
  \href{http://dx.doi.org/10.1046/j.1365-8711.2003.06333.x}{\JournalTitle{\mnras},
  340, 1095}

\bibitem[{{Elvis} {et~al.}(1994){Elvis}, {Wilkes}, {McDowell}, {Green},
  {Bechtold}, {Willner}, {Oey}, {Polomski}, \& {Cutri}}]{Elvis1994}
{Elvis}, M., {Wilkes}, B.~J., {McDowell}, J.~C., {et~al.} 1994,
  \href{http://dx.doi.org/10.1086/192093}{\JournalTitle{\apjs}, 95, 1}

\bibitem[{{Flewelling} {et~al.}(2020){Flewelling}, {Magnier}, {Chambers},
  {Heasley}, {Holmberg}, {Huber}, {Sweeney}, {Waters}, {Calamida}, {Casertano},
  {Chen}, {Farrow}, {Hasinger}, {Henderson}, {Long}, {Metcalfe}, {Narayan},
  {Nieto-Santisteban}, {Norberg}, {Rest}, {Saglia}, {Szalay}, {Thakar},
  {Tonry}, {Valenti}, {Werner}, {White}, {Denneau}, {Draper}, {Hodapp},
  {Jedicke}, {Kaiser}, {Kudritzki}, {Price}, {Wainscoat}, {Chastel}, {McLean},
  {Postman}, \& {Shiao}}]{Flewellin2020}
{Flewelling}, H.~A., {Magnier}, E.~A., {Chambers}, K.~C., {et~al.} 2020,
  \href{http://dx.doi.org/10.3847/1538-4365/abb82d}{\JournalTitle{\apjs}, 251,
  7}

\bibitem[{{Floyd} {et~al.}(2013){Floyd}, {Dunlop}, {Kukula}, {Brown}, {McLure},
  {Baum}, \& {O'Dea}}]{2013MNRAS.429....2F}
{Floyd}, D. J.~E., {Dunlop}, J.~S., {Kukula}, M.~J., {et~al.} 2013,
  \href{http://dx.doi.org/10.1093/mnras/sts291}{\JournalTitle{\mnras}, 429, 2}

\bibitem[{{Floyd} {et~al.}(2004){Floyd}, {Kukula}, {Dunlop}, {McLure},
  {Miller}, {Percival}, {Baum}, \& {O'Dea}}]{2004MNRAS.355..196F}
{Floyd}, D. J.~E., {Kukula}, M.~J., {Dunlop}, J.~S., {et~al.} 2004,
  \href{http://dx.doi.org/10.1111/j.1365-2966.2004.08315.x}{\JournalTitle{\mnras},
  355, 196}

\bibitem[{{Gaia Collaboration} {et~al.}(2023){Gaia Collaboration}, {Vallenari},
  {Brown}, {Prusti}, {de Bruijne}, {Arenou}, {Babusiaux}, {Biermann},
  {Creevey}, {Ducourant}, {Evans}, {Eyer}, {Guerra}, {Hutton}, {Jordi},
  {Klioner}, {Lammers}, {Lindegren}, {Luri}, {Mignard}, {Panem}, {Pourbaix},
  {Randich}, {Sartoretti}, {Soubiran}, {Tanga}, {Walton}, {Bailer-Jones},
  {Bastian}, {Drimmel}, {Jansen}, {Katz}, {Lattanzi}, {van Leeuwen}, {Bakker},
  {Cacciari}, {Casta{\~n}eda}, {De Angeli}, {Fabricius}, {Fouesneau},
  {Fr{\'e}mat}, {Galluccio}, {Guerrier}, {Heiter}, {Masana}, {Messineo},
  {Mowlavi}, {Nicolas}, {Nienartowicz}, {Pailler}, {Panuzzo}, {Riclet}, {Roux},
  {Seabroke}, {Sordo}, {Th{\'e}venin}, {Gracia-Abril}, {Portell}, {Teyssier},
  {Altmann}, {Andrae}, {Audard}, {Bellas-Velidis}, {Benson}, {Berthier},
  {Blomme}, {Burgess}, {Busonero}, {Busso}, {C{\'a}novas}, {Carry}, {Cellino},
  {Cheek}, {Clementini}, {Damerdji}, {Davidson}, {de Teodoro}, {Nu{\~n}ez
  Campos}, {Delchambre}, {Dell'Oro}, {Esquej}, {Fern{\'a}ndez-Hern{\'a}ndez},
  {Fraile}, {Garabato}, {Garc{\'\i}a-Lario}, {Gosset}, {Haigron}, {Halbwachs},
  {Hambly}, {Harrison}, {Hern{\'a}ndez}, {Hestroffer}, {Hodgkin}, {Holl},
  {Jan{\ss}en}, {Jevardat de Fombelle}, {Jordan}, {Krone-Martins}, {Lanzafame},
  {L{\"o}ffler}, {Marchal}, {Marrese}, {Moitinho}, {Muinonen}, {Osborne},
  {Pancino}, {Pauwels}, {Recio-Blanco}, {Reyl{\'e}}, {Riello}, {Rimoldini},
  {Roegiers}, {Rybizki}, {Sarro}, {Siopis}, {Smith}, {Sozzetti}, {Utrilla},
  {van Leeuwen}, {Abbas}, {{\'A}brah{\'a}m}, {Abreu Aramburu}, {Aerts},
  {Aguado}, {Ajaj}, {Aldea-Montero}, {Altavilla}, {{\'A}lvarez}, {Alves},
  {Anders}, {Anderson}, {Anglada Varela}, {Antoja}, {Baines}, {Baker},
  {Balaguer-N{\'u}{\~n}ez}, {Balbinot}, {Balog}, {Barache}, {Barbato},
  {Barros}, {Barstow}, {Bartolom{\'e}}, {Bassilana}, {Bauchet}, {Becciani},
  {Bellazzini}, {Berihuete}, {Bernet}, {Bertone}, {Bianchi}, {Binnenfeld},
  {Blanco-Cuaresma}, {Blazere}, {Boch}, {Bombrun}, {Bossini}, {Bouquillon},
  {Bragaglia}, {Bramante}, {Breedt}, {Bressan}, {Brouillet}, {Brugaletta},
  {Bucciarelli}, {Burlacu}, {Butkevich}, {Buzzi}, {Caffau}, {Cancelliere},
  {Cantat-Gaudin}, {Carballo}, {Carlucci}, {Carnerero}, {Carrasco},
  {Casamiquela}, {Castellani}, {Castro-Ginard}, {Chaoul}, {Charlot}, {Chemin},
  {Chiaramida}, {Chiavassa}, {Chornay}, {Comoretto}, {Contursi}, {Cooper},
  {Cornez}, {Cowell}, {Crifo}, {Cropper}, {Crosta}, {Crowley}, {Dafonte},
  {Dapergolas}, {David}, {David}, {de Laverny}, {De Luise}, {De March}, {De
  Ridder}, {de Souza}, {de Torres}, {del Peloso}, {del Pozo}, {Delbo},
  {Delgado}, {Delisle}, {Demouchy}, {Dharmawardena}, {Di Matteo}, {Diakite},
  {Diener}, {Distefano}, {Dolding}, {Edvardsson}, {Enke}, {Fabre}, {Fabrizio},
  {Faigler}, {Fedorets}, {Fernique}, {Fienga}, {Figueras}, {Fournier},
  {Fouron}, {Fragkoudi}, {Gai}, {Garcia-Gutierrez}, {Garcia-Reinaldos},
  {Garc{\'\i}a-Torres}, {Garofalo}, {Gavel}, {Gavras}, {Gerlach}, {Geyer},
  {Giacobbe}, {Gilmore}, {Girona}, {Giuffrida}, {Gomel}, {Gomez},
  {Gonz{\'a}lez-N{\'u}{\~n}ez}, {Gonz{\'a}lez-Santamar{\'\i}a},
  {Gonz{\'a}lez-Vidal}, {Granvik}, {Guillout}, {Guiraud},
  {Guti{\'e}rrez-S{\'a}nchez}, {Guy}, {Hatzidimitriou}, {Hauser}, {Haywood},
  {Helmer}, {Helmi}, {Sarmiento}, {Hidalgo}, {Hilger}, {H{\l}adczuk}, {Hobbs},
  {Holland}, {Huckle}, {Jardine}, {Jasniewicz}, {Jean-Antoine Piccolo},
  {Jim{\'e}nez-Arranz}, {Jorissen}, {Juaristi Campillo}, {Julbe}, {Karbevska},
  {Kervella}, {Khanna}, {Kontizas}, {Kordopatis}, {Korn}, {K{\'o}sp{\'a}l},
  {Kostrzewa-Rutkowska}, {Kruszy{\'n}ska}, {Kun}, {Laizeau}, {Lambert},
  {Lanza}, {Lasne}, {Le Campion}, {Lebreton}, {Lebzelter}, {Leccia}, {Leclerc},
  {Lecoeur-Taibi}, {Liao}, {Licata}, {Lindstr{\o}m}, {Lister}, {Livanou},
  {Lobel}, {Lorca}, {Loup}, {Madrero Pardo}, {Magdaleno Romeo}, {Managau},
  {Mann}, {Manteiga}, {Marchant}, {Marconi}, {Marcos}, {Marcos Santos},
  {Mar{\'\i}n Pina}, {Marinoni}, {Marocco}, {Marshall}, {Martin Polo},
  {Mart{\'\i}n-Fleitas}, {Marton}, {Mary}, {Masip}, {Massari},
  {Mastrobuono-Battisti}, {Mazeh}, {McMillan}, {Messina}, {Michalik}, {Millar},
  {Mints}, {Molina}, {Molinaro}, {Moln{\'a}r}, {Monari}, {Mongui{\'o}},
  {Montegriffo}, {Montero}, {Mor}, {Mora}, {Morbidelli}, {Morel}, {Morris},
  {Muraveva}, {Murphy}, {Musella}, {Nagy}, {Noval}, {Oca{\~n}a}, {Ogden},
  {Ordenovic}, {Osinde}, {Pagani}, {Pagano}, {Palaversa}, {Palicio},
  {Pallas-Quintela}, {Panahi}, {Payne-Wardenaar}, {Pe{\~n}alosa Esteller},
  {Penttil{\"a}}, {Pichon}, {Piersimoni}, {Pineau}, {Plachy}, {Plum}, {Poggio},
  {Pr{\v{s}}a}, {Pulone}, {Racero}, {Ragaini}, {Rainer}, {Raiteri}, {Rambaux},
  {Ramos}, {Ramos-Lerate}, {Re Fiorentin}, {Regibo}, {Richards}, {Rios Diaz},
  {Ripepi}, {Riva}, {Rix}, {Rixon}, {Robichon}, {Robin}, {Robin}, {Roelens},
  {Rogues}, {Rohrbasser}, {Romero-G{\'o}mez}, {Rowell}, {Royer}, {Ruz Mieres},
  {Rybicki}, {Sadowski}, {S{\'a}ez N{\'u}{\~n}ez}, {Sagrist{\`a} Sell{\'e}s},
  {Sahlmann}, {Salguero}, {Samaras}, {Sanchez Gimenez}, {Sanna},
  {Santove{\~n}a}, {Sarasso}, {Schultheis}, {Sciacca}, {Segol}, {Segovia},
  {S{\'e}gransan}, {Semeux}, {Shahaf}, {Siddiqui}, {Siebert}, {Siltala},
  {Silvelo}, {Slezak}, {Slezak}, {Smart}, {Snaith}, {Solano}, {Solitro},
  {Souami}, {Souchay}, {Spagna}, {Spina}, {Spoto}, {Steele},
  {Steidelm{\"u}ller}, {Stephenson}, {S{\"u}veges}, {Surdej}, {Szabados},
  {Szegedi-Elek}, {Taris}, {Taylor}, {Teixeira}, {Tolomei}, {Tonello}, {Torra},
  {Torra}, {Torralba Elipe}, {Trabucchi}, {Tsounis}, {Turon}, {Ulla}, {Unger},
  {Vaillant}, {van Dillen}, {van Reeven}, {Vanel}, {Vecchiato}, {Viala},
  {Vicente}, {Voutsinas}, {Weiler}, {Wevers}, {Wyrzykowski}, {Yoldas}, {Yvard},
  {Zhao}, {Zorec}, {Zucker}, \& {Zwitter}}]{GaiaDR3}
{Gaia Collaboration}, {Vallenari}, A., {Brown}, A.~G.~A., {et~al.} 2023,
  \href{http://dx.doi.org/10.1051/0004-6361/202243940}{\JournalTitle{\aap},
  674, A1}

\bibitem[{{Gehrels}(1986)}]{Gehrels1986}
{Gehrels}, N. 1986,
  \href{http://dx.doi.org/10.1086/164079}{\JournalTitle{\apj}, 303, 336}

\bibitem[{{Gordon} {et~al.}(2021){Gordon}, {Boyce}, {O'Dea}, {Rudnick},
  {Andernach}, {Vantyghem}, {Baum}, {Bui}, {Dionyssiou}, {Safi-Harb}, \&
  {Sander}}]{Gordon2021}
{Gordon}, Y.~A., {Boyce}, M.~M., {O'Dea}, C.~P., {et~al.} 2021,
  \href{http://dx.doi.org/10.3847/1538-4365/ac05c0}{\JournalTitle{\apjs}, 255,
  30}

\bibitem[{{Green} {et~al.}(2022){Green}, {Pulgarin-Duque}, {Anderson},
  {MacLeod}, {Eracleous}, {Ruan}, {Runnoe}, {Graham}, {Roulston}, {Schneider},
  {Ahlf}, {Bizyaev}, {Brownstein}, {del Casal}, {Dodd}, {Hoover}, {Matt},
  {Merloni}, {Pan}, {Ramirez}, {Ridder}, \& {Moseley}}]{Green2022}
{Green}, P.~J., {Pulgarin-Duque}, L., {Anderson}, S.~F., {et~al.} 2022,
  \href{http://dx.doi.org/10.3847/1538-4357/ac743f}{\JournalTitle{\apj}, 933,
  180}

\bibitem[{{Guo} {et~al.}(2017{\natexlab{a}}){Guo}, {Wang}, {Cai}, \&
  {Sun}}]{Guo2017ApJ...847..132G}
{Guo}, H., {Wang}, J., {Cai}, Z., \& {Sun}, M. 2017{\natexlab{a}},
  \href{http://dx.doi.org/10.3847/1538-4357/aa8d71}{\JournalTitle{\apj}, 847,
  132}

\bibitem[{{Guo} {et~al.}(2017{\natexlab{b}}){Guo}, {Wang}, {Cai}, \&
  {Sun}}]{Guo2017}
---. 2017{\natexlab{b}},
  \href{http://dx.doi.org/10.3847/1538-4357/aa8d71}{\JournalTitle{\apj}, 847,
  132}

\bibitem[{{Gupta} \& {Joshi}(2005)}]{Gupta2005}
{Gupta}, A.~C., \& {Joshi}, U.~C. 2005,
  \href{http://dx.doi.org/10.1051/0004-6361:20042370}{\JournalTitle{\aap}, 440,
  855}

\bibitem[{Helfand {et~al.}(2001)Helfand, Stone, Willman, White, Becker, Price,
  Gregg, \& McMahon}]{helfand2001long}
Helfand, D.~J., Stone, R.~P., Willman, B., {et~al.} 2001, \JournalTitle{The
  Astronomical Journal}, 121, 1872

\bibitem[{{Ho}(2002)}]{Ho2002}
{Ho}, L.~C. 2002, \href{http://dx.doi.org/10.1086/324399}{\JournalTitle{\apj},
  564, 120}

\bibitem[{{Hu} {et~al.}(2023){Hu}, {Cai}, \& {Wang}}]{HuXF2023}
{Hu}, X.-F., {Cai}, Z.-Y., \& {Wang}, J.-X. 2023,
  \href{http://dx.doi.org/10.48550/arXiv.2310.16223}{\JournalTitle{arXiv
  e-prints}, arXiv:2310.16223}

\bibitem[{{Isobe} {et~al.}(1990){Isobe}, {Feigelson}, {Akritas}, \&
  {Babu}}]{1990ApJ...364..104I}
{Isobe}, T., {Feigelson}, E.~D., {Akritas}, M.~G., \& {Babu}, G.~J. 1990,
  \href{http://dx.doi.org/10.1086/169390}{\JournalTitle{\apj}, 364, 104}

\bibitem[{{Jiang} {et~al.}(2007){Jiang}, {Fan}, {Ivezi{\'c}}, {Richards},
  {Schneider}, {Strauss}, \& {Kelly}}]{Jiang2007ApJ...656..680J}
{Jiang}, L., {Fan}, X., {Ivezi{\'c}}, {\v{Z}}., {et~al.} 2007,
  \href{http://dx.doi.org/10.1086/510831}{\JournalTitle{\apj}, 656, 680}

\bibitem[{{Kang} {et~al.}(2018){Kang}, {Wang}, {Cai}, {Guo}, {Zhu}, {Cao},
  {Gu}, \& {Yuan}}]{2018ApJ...868...58K}
{Kang}, W.-Y., {Wang}, J.-X., {Cai}, Z.-Y., {et~al.} 2018,
  \href{http://dx.doi.org/10.3847/1538-4357/aae6c4}{\JournalTitle{\apj}, 868,
  58}

\bibitem[{{Kang} {et~al.}(2021){Kang}, {Wang}, {Cai}, \&
  {Ren}}]{2021ApJ...911..148K}
{Kang}, W.-Y., {Wang}, J.-X., {Cai}, Z.-Y., \& {Ren}, W.-K. 2021,
  \href{http://dx.doi.org/10.3847/1538-4357/abeb69}{\JournalTitle{\apj}, 911,
  148}

\bibitem[{{Kellermann} {et~al.}(2016){Kellermann}, {Condon}, {Kimball},
  {Perley}, \& {Ivezi{\'c}}}]{2016ApJ...831..168K}
{Kellermann}, K.~I., {Condon}, J.~J., {Kimball}, A.~E., {Perley}, R.~A., \&
  {Ivezi{\'c}}, {\v{Z}}. 2016,
  \href{http://dx.doi.org/10.3847/0004-637X/831/2/168}{\JournalTitle{\apj},
  831, 168}

\bibitem[{{Kellermann} {et~al.}(1989){Kellermann}, {Sramek}, {Schmidt},
  {Shaffer}, \& {Green}}]{1989AJ.....98.1195K}
{Kellermann}, K.~I., {Sramek}, R., {Schmidt}, M., {Shaffer}, D.~B., \& {Green},
  R. 1989, \href{http://dx.doi.org/10.1086/115207}{\JournalTitle{\aj}, 98,
  1195}

\bibitem[{{Kelly} {et~al.}(2009){Kelly}, {Bechtold}, \&
  {Siemiginowska}}]{Kelly2009}
{Kelly}, B.~C., {Bechtold}, J., \& {Siemiginowska}, A. 2009,
  \href{http://dx.doi.org/10.1088/0004-637X/698/1/895}{\JournalTitle{\apj},
  698, 895}

\bibitem[{{Koz{\l}owski}(2016)}]{Kozlowski2016ApJ826}
{Koz{\l}owski}, S. 2016,
  \href{http://dx.doi.org/10.3847/0004-637X/826/2/118}{\JournalTitle{\apj},
  826, 118}

\bibitem[{{Lacy} {et~al.}(2001){Lacy}, {Laurent-Muehleisen}, {Ridgway},
  {Becker}, \& {White}}]{Lacy2001}
{Lacy}, M., {Laurent-Muehleisen}, S.~A., {Ridgway}, S.~E., {Becker}, R.~H., \&
  {White}, R.~L. 2001,
  \href{http://dx.doi.org/10.1086/319836}{\JournalTitle{\apjl}, 551, L17}

\bibitem[{{LaMassa} {et~al.}(2015){LaMassa}, {Cales}, {Moran}, {Myers},
  {Richards}, {Eracleous}, {Heckman}, {Gallo}, \& {Urry}}]{LaMassa2015}
{LaMassa}, S.~M., {Cales}, S., {Moran}, E.~C., {et~al.} 2015,
  \href{http://dx.doi.org/10.1088/0004-637X/800/2/144}{\JournalTitle{\apj},
  800, 144}

\bibitem[{{Laor}(2000)}]{Laor2000}
{Laor}, A. 2000, \href{http://dx.doi.org/10.1086/317280}{\JournalTitle{\apjl},
  543, L111}

\bibitem[{{Li} \& {Begelman}(2014)}]{Li2014}
{Li}, S.-L., \& {Begelman}, M.~C. 2014,
  \href{http://dx.doi.org/10.1088/0004-637X/786/1/6}{\JournalTitle{\apj}, 786,
  6}

\bibitem[{{Liao} {et~al.}(2022){Liao}, {Wang}, {Kang}, \& {Zhou}}]{Liao2022}
{Liao}, M., {Wang}, J., {Kang}, W., \& {Zhou}, M. 2022,
  \href{http://dx.doi.org/10.1093/mnras/stac266}{\JournalTitle{\mnras}, 512,
  296}

\bibitem[{{Livio} {et~al.}(1999){Livio}, {Ogilvie}, \& {Pringle}}]{Livio1999}
{Livio}, M., {Ogilvie}, G.~I., \& {Pringle}, J.~E. 1999,
  \href{http://dx.doi.org/10.1086/306777}{\JournalTitle{\apj}, 512, 100}

\bibitem[{{Lobanov}(1998)}]{Lobanov1998}
{Lobanov}, A.~P. 1998,
  \href{http://dx.doi.org/10.48550/arXiv.astro-ph/9712132}{\JournalTitle{\aap},
  330, 79}

\bibitem[{{Lyke} {et~al.}(2020){Lyke}, {Higley}, {McLane}, {Schurhammer},
  {Myers}, {Ross}, {Dawson}, {Chabanier}, {Martini}, {Busca}, {Mas des
  Bourboux}, {Salvato}, {Streblyanska}, {Zarrouk}, {Burtin}, {Anderson},
  {Bautista}, {Bizyaev}, {Brandt}, {Brinkmann}, {Brownstein}, {Comparat},
  {Green}, {de la Macorra}, {Mu{\~n}oz Guti{\'e}rrez}, {Hou}, {Newman},
  {Palanque-Delabrouille}, {P{\^a}ris}, {Percival}, {Petitjean}, {Rich},
  {Rossi}, {Schneider}, {Smith}, {Vivek}, \& {Weaver}}]{Lyke2020}
{Lyke}, B.~W., {Higley}, A.~N., {McLane}, J.~N., {et~al.} 2020,
  \href{http://dx.doi.org/10.3847/1538-4365/aba623}{\JournalTitle{\apjs}, 250,
  8}

\bibitem[{{MacLeod} {et~al.}(2010){MacLeod}, {Ivezi{\'c}}, {Kochanek},
  {Koz{\l}owski}, {Kelly}, {Bullock}, {Kimball}, {Sesar}, {Westman}, {Brooks},
  {Gibson}, {Becker}, \& {de Vries}}]{Macleod2010}
{MacLeod}, C.~L., {Ivezi{\'c}}, {\v Z}., {Kochanek}, C.~S., {et~al.} 2010,
  \href{http://dx.doi.org/10.1088/0004-637X/721/2/1014}{\JournalTitle{\apj},
  721, 1014}

\bibitem[{{MacLeod} {et~al.}(2012){MacLeod}, {Ivezi{\'c}}, {Sesar}, {de Vries},
  {Kochanek}, {Kelly}, {Becker}, {Lupton}, {Hall}, {Richards}, {Anderson}, \&
  {Schneider}}]{Macleod2012}
{MacLeod}, C.~L., {Ivezi{\'c}}, {\v Z}., {Sesar}, B., {et~al.} 2012,
  \href{http://dx.doi.org/10.1088/0004-637X/753/2/106}{\JournalTitle{\apj},
  753, 106}

\bibitem[{{MacLeod} {et~al.}(2016){MacLeod}, {Ross}, {Lawrence}, {Goad},
  {Horne}, {Burgett}, {Chambers}, {Flewelling}, {Hodapp}, {Kaiser}, {Magnier},
  {Wainscoat}, \& {Waters}}]{Macleod2016}
{MacLeod}, C.~L., {Ross}, N.~P., {Lawrence}, A., {et~al.} 2016,
  \href{http://dx.doi.org/10.1093/mnras/stv2997}{\JournalTitle{\mnras}, 457,
  389}

\bibitem[{{MacLeod} {et~al.}(2019){MacLeod}, {Green}, {Anderson}, {Bruce},
  {Eracleous}, {Graham}, {Homan}, {Lawrence}, {LeBleu}, {Ross}, {Ruan},
  {Runnoe}, {Stern}, {Burgett}, {Chambers}, {Kaiser}, {Magnier}, \&
  {Metcalfe}}]{2019ApJ...874....8M}
{MacLeod}, C.~L., {Green}, P.~J., {Anderson}, S.~F., {et~al.} 2019,
  \href{http://dx.doi.org/10.3847/1538-4357/ab05e2}{\JournalTitle{\apj}, 874,
  8}

\bibitem[{{Meusinger} {et~al.}(2011){Meusinger}, {Hinze}, \& {de
  Hoon}}]{2011A&A...525A..37M}
{Meusinger}, H., {Hinze}, A., \& {de Hoon}, A. 2011,
  \href{http://dx.doi.org/10.1051/0004-6361/201015520}{\JournalTitle{\aap},
  525, A37}

\bibitem[{{Meusinger} \& {Weiss}(2013)}]{2013A&A...560A.104M}
{Meusinger}, H., \& {Weiss}, V. 2013,
  \href{http://dx.doi.org/10.1051/0004-6361/201322410}{\JournalTitle{\aap},
  560, A104}

\bibitem[{{Miller} {et~al.}(1993){Miller}, {Rawlings}, \&
  {Saunders}}]{1993MNRAS.263..425M}
{Miller}, P., {Rawlings}, S., \& {Saunders}, R. 1993,
  \href{http://dx.doi.org/10.1093/mnras/263.2.425}{\JournalTitle{\mnras}, 263,
  425}

\bibitem[{{Padovani} {et~al.}(2017){Padovani}, {Alexander}, {Assef}, {De
  Marco}, {Giommi}, {Hickox}, {Richards}, {Smol{\v{c}}i{\'c}},
  {Hatziminaoglou}, {Mainieri}, \& {Salvato}}]{2017A&ARv..25....2P}
{Padovani}, P., {Alexander}, D.~M., {Assef}, R.~J., {et~al.} 2017,
  \href{http://dx.doi.org/10.1007/s00159-017-0102-9}{\JournalTitle{\aapr}, 25,
  2}

\bibitem[{{Panessa} {et~al.}(2019){Panessa}, {Baldi}, {Laor}, {Padovani},
  {Behar}, \& {McHardy}}]{2019NatAs...3..387P}
{Panessa}, F., {Baldi}, R.~D., {Laor}, A., {et~al.} 2019,
  \href{http://dx.doi.org/10.1038/s41550-019-0765-4}{\JournalTitle{Nature
  Astronomy}, 3, 387}

\bibitem[{{Pessah} \& {Psaltis}(2005)}]{Pessah2005}
{Pessah}, M.~E., \& {Psaltis}, D. 2005,
  \href{http://dx.doi.org/10.1086/430940}{\JournalTitle{\apj}, 628, 879}

\bibitem[{{Reines} {et~al.}(2020){Reines}, {Condon}, {Darling}, \&
  {Greene}}]{2020ApJ...888...36R}
{Reines}, A.~E., {Condon}, J.~J., {Darling}, J., \& {Greene}, J.~E. 2020,
  \href{http://dx.doi.org/10.3847/1538-4357/ab4999}{\JournalTitle{\apj}, 888,
  36}

\bibitem[{{Ren} {et~al.}(2022){Ren}, {Wang}, {Cai}, \&
  {Guo}}]{2022ApJ...925...50R}
{Ren}, W., {Wang}, J., {Cai}, Z., \& {Guo}, H. 2022,
  \href{http://dx.doi.org/10.3847/1538-4357/ac3828}{\JournalTitle{\apj}, 925,
  50}

\bibitem[{{Ricci} \& {Trakhtenbrot}(2022)}]{Ricci2022}
{Ricci}, C., \& {Trakhtenbrot}, B. 2022,
  \href{http://dx.doi.org/10.48550/arXiv.2211.05132}{\JournalTitle{arXiv
  e-prints}, arXiv:2211.05132}

\bibitem[{{Rumbaugh} {et~al.}(2018){Rumbaugh}, {Shen}, {Morganson}, {Liu},
  {Banerji}, {McMahon}, {Abdalla}, {Benoit-L{\'e}vy}, {Bertin}, {Brooks},
  {Buckley-Geer}, {Capozzi}, {Carnero Rosell}, {Carrasco Kind}, {Carretero},
  {Cunha}, {D'Andrea}, {da Costa}, {DePoy}, {Desai}, {Doel}, {Frieman},
  {Garc{\'\i}a-Bellido}, {Gruen}, {Gruendl}, {Gschwend}, {Gutierrez},
  {Honscheid}, {James}, {Kuehn}, {Kuhlmann}, {Kuropatkin}, {Lima}, {Maia},
  {Marshall}, {Martini}, {Menanteau}, {Plazas}, {Reil}, {Roodman}, {Sanchez},
  {Scarpine}, {Schindler}, {Schubnell}, {Sheldon}, {Smith}, {Soares-Santos},
  {Sobreira}, {Suchyta}, {Swanson}, {Walker}, {Wester}, \& {DES
  Collaboration}}]{2018ApJ...854..160R}
{Rumbaugh}, N., {Shen}, Y., {Morganson}, E., {et~al.} 2018,
  \href{http://dx.doi.org/10.3847/1538-4357/aaa9b6}{\JournalTitle{\apj}, 854,
  160}

\bibitem[{{S{{a}}dowski}(2016)}]{Sadowski2016}
{S{{a}}dowski}, A. 2016,
  \href{http://dx.doi.org/10.1093/mnras/stw913}{\JournalTitle{\mnras}, 459,
  4397}

\bibitem[{{Sesar} {et~al.}(2007){Sesar}, {Ivezi{\'c}}, {Lupton}, {Juri{\'c}},
  {Gunn}, {Knapp}, {DeLee}, {Smith}, {Miknaitis}, {Lin}, {Tucker}, {Doi},
  {Tanaka}, {Fukugita}, {Holtzman}, {Kent}, {Yanny}, {Schlegel}, {Finkbeiner},
  {Padmanabhan}, {Rockosi}, {Bond}, {Lee}, {Stoughton}, {Jester}, {Harris},
  {Harding}, {Brinkmann}, {Schneider}, {York}, {Richmond}, \& {Vanden
  Berk}}]{Sesar2007}
{Sesar}, B., {Ivezi{\'c}}, {\v Z}., {Lupton}, R.~H., {et~al.} 2007,
  \href{http://dx.doi.org/10.1086/521819}{\JournalTitle{\aj}, 134, 2236}

\bibitem[{{Shang} {et~al.}(2011){Shang}, {Brotherton}, {Wills}, {Wills},
  {Cales}, {Dale}, {Green}, {Runnoe}, {Nemmen}, {Gallagher}, {Ganguly},
  {Hines}, {Kelly}, {Kriss}, {Li}, {Tang}, \& {Xie}}]{Shang2011}
{Shang}, Z., {Brotherton}, M.~S., {Wills}, B.~J., {et~al.} 2011,
  \href{http://dx.doi.org/10.1088/0067-0049/196/1/2}{\JournalTitle{\apjs}, 196,
  2}

\bibitem[{{Sheng} {et~al.}(2017){Sheng}, {Wang}, {Jiang}, {Yang}, {Yan}, {Dou},
  \& {Peng}}]{Sheng2017}
{Sheng}, Z., {Wang}, T., {Jiang}, N., {et~al.} 2017,
  \href{http://dx.doi.org/10.3847/2041-8213/aa85de}{\JournalTitle{\apjl}, 846,
  L7}

\bibitem[{{Sheng} {et~al.}(2020){Sheng}, {Wang}, {Jiang}, {Ding}, {Cai}, {Guo},
  {Sun}, {Dou}, \& {Yang}}]{Sheng2020ApJ...889...46S}
---. 2020,
  \href{http://dx.doi.org/10.3847/1538-4357/ab5af9}{\JournalTitle{\apj}, 889,
  46}

\bibitem[{{Sikora} {et~al.}(2007){Sikora}, {Stawarz}, \& {Lasota}}]{Sikora2007}
{Sikora}, M., {Stawarz}, {\L}., \& {Lasota}, J.-P. 2007,
  \href{http://dx.doi.org/10.1086/511972}{\JournalTitle{\apj}, 658, 815}

\bibitem[{{Stoughton} {et~al.}(2002){Stoughton}, {Lupton}, {Bernardi},
  {Blanton}, {Burles}, {Castander}, {Connolly}, {Eisenstein}, {Frieman},
  {Hennessy}, {Hindsley}, {Ivezi{\'c}}, {Kent}, {Kunszt}, {Lee}, {Meiksin},
  {Munn}, {Newberg}, {Nichol}, {Nicinski}, {Pier}, {Richards}, {Richmond},
  {Schlegel}, {Smith}, {Strauss}, {SubbaRao}, {Szalay}, {Thakar}, {Tucker},
  {Vanden Berk}, {Yanny}, {Adelman}, {Anderson}, {Anderson}, {Annis},
  {Bahcall}, {Bakken}, {Bartelmann}, {Bastian}, {Bauer}, {Berman},
  {B{\"o}hringer}, {Boroski}, {Bracker}, {Briegel}, {Briggs}, {Brinkmann},
  {Brunner}, {Carey}, {Carr}, {Chen}, {Christian}, {Colestock}, {Crocker},
  {Csabai}, {Czarapata}, {Dalcanton}, {Davidsen}, {Davis}, {Dehnen},
  {Dodelson}, {Doi}, {Dombeck}, {Donahue}, {Ellman}, {Elms}, {Evans}, {Eyer},
  {Fan}, {Federwitz}, {Friedman}, {Fukugita}, {Gal}, {Gillespie}, {Glazebrook},
  {Gray}, {Grebel}, {Greenawalt}, {Greene}, {Gunn}, {de Haas}, {Haiman},
  {Haldeman}, {Hall}, {Hamabe}, {Hansen}, {Harris}, {Harris}, {Harvanek},
  {Hawley}, {Hayes}, {Heckman}, {Helmi}, {Henden}, {Hogan}, {Hogg}, {Holmgren},
  {Holtzman}, {Huang}, {Hull}, {Ichikawa}, {Ichikawa}, {Johnston}, {Kauffmann},
  {Kim}, {Kimball}, {Kinney}, {Klaene}, {Kleinman}, {Klypin}, {Knapp},
  {Korienek}, {Krolik}, {Kron}, {Krzesi{\'n}ski}, {Lamb}, {Leger},
  {Limmongkol}, {Lindenmeyer}, {Long}, {Loomis}, {Loveday}, {MacKinnon},
  {Mannery}, {Mantsch}, {Margon}, {McGehee}, {McKay}, {McLean}, {Menou},
  {Merelli}, {Mo}, {Monet}, {Nakamura}, {Narayanan}, {Nash}, {Neilsen},
  {Newman}, {Nitta}, {Odenkirchen}, {Okada}, {Okamura}, {Ostriker}, {Owen},
  {Pauls}, {Peoples}, {Peterson}, {Petravick}, {Pope}, {Pordes}, {Postman},
  {Prosapio}, {Quinn}, {Rechenmacher}, {Rivetta}, {Rix}, {Rockosi}, {Rosner},
  {Ruthmansdorfer}, {Sandford}, {Schneider}, {Scranton}, {Sekiguchi}, {Sergey},
  {Sheth}, {Shimasaku}, {Smee}, {Snedden}, {Stebbins}, {Stubbs}, {Szapudi},
  {Szkody}, {Szokoly}, {Tabachnik}, {Tsvetanov}, {Uomoto}, {Vogeley}, {Voges},
  {Waddell}, {Walterbos}, {Wang}, {Watanabe}, {Weinberg}, {White}, {White},
  {Wilhite}, {Wolfe}, {Yasuda}, {York}, {Zehavi}, \& {Zheng}}]{Stoughton2002}
{Stoughton}, C., {Lupton}, R.~H., {Bernardi}, M., {et~al.} 2002,
  \href{http://dx.doi.org/10.1086/324741}{\JournalTitle{\aj}, 123, 485}

\bibitem[{{Sun} {et~al.}(2018){Sun}, {Xue}, {Wang}, {Cai}, \&
  {Guo}}]{2018ApJ...866...74S}
{Sun}, M., {Xue}, Y., {Wang}, J., {Cai}, Z., \& {Guo}, H. 2018,
  \href{http://dx.doi.org/10.3847/1538-4357/aae208}{\JournalTitle{\apj}, 866,
  74}

\bibitem[{{Sun} {et~al.}(2014){Sun}, {Wang}, {Chen}, \& {Zheng}}]{Sun2014}
{Sun}, Y.-H., {Wang}, J.-X., {Chen}, X.-Y., \& {Zheng}, Z.-Y. 2014,
  \href{http://dx.doi.org/10.1088/0004-637X/792/1/54}{\JournalTitle{\apj}, 792,
  54}

\bibitem[{{Timmer} \& {Koenig}(1995)}]{Timmer1995A&A...300..707T}
{Timmer}, J., \& {Koenig}, M. 1995, \JournalTitle{\aap}, 300, 707

\bibitem[{{Ulrich} {et~al.}(1997){Ulrich}, {Maraschi}, \& {Urry}}]{Ulrich1997}
{Ulrich}, M.-H., {Maraschi}, L., \& {Urry}, C.~M. 1997,
  \href{http://dx.doi.org/10.1146/annurev.astro.35.1.445}{\JournalTitle{\araa},
  35, 445}

\bibitem[{{Vanden Berk} {et~al.}(2004){Vanden Berk}, {Wilhite}, {Kron},
  {Anderson}, {Brunner}, {Hall}, {Ivezi{\'c}}, {Richards}, {Schneider}, {York},
  {Brinkmann}, {Lamb}, {Nichol}, \& {Schlegel}}]{berk2004ensemble}
{Vanden Berk}, D.~E., {Wilhite}, B.~C., {Kron}, R.~G., {et~al.} 2004,
  \href{http://dx.doi.org/10.1086/380563}{\JournalTitle{\apj}, 601, 692}

\bibitem[{{Vaughan} {et~al.}(2003){Vaughan}, {Edelson}, {Warwick}, \&
  {Uttley}}]{2003MNRAS.345.1271V}
{Vaughan}, S., {Edelson}, R., {Warwick}, R.~S., \& {Uttley}, P. 2003,
  \href{http://dx.doi.org/10.1046/j.1365-2966.2003.07042.x}{\JournalTitle{\mnras},
  345, 1271}

\bibitem[{{Webb} {et~al.}(2020){Webb}, {Coriat}, {Traulsen}, {Ballet}, {Motch},
  {Carrera}, {Koliopanos}, {Authier}, {de la Calle}, {Ceballos}, {Colomo},
  {Chuard}, {Freyberg}, {Garcia}, {Kolehmainen}, {Lamer}, {Lin}, {Maggi},
  {Michel}, {Page}, {Page}, {Perea-Calderon}, {Pineau}, {Rodriguez}, {Rosen},
  {Santos Lleo}, {Saxton}, {Schwope}, {Tom{\'a}s}, {Watson}, \&
  {Zakardjian}}]{2020A&A...641A.136W}
{Webb}, N.~A., {Coriat}, M., {Traulsen}, I., {et~al.} 2020,
  \href{http://dx.doi.org/10.1051/0004-6361/201937353}{\JournalTitle{\aap},
  641, A136}

\bibitem[{{White} {et~al.}(1997){White}, {Becker}, {Helfand}, \&
  {Gregg}}]{1997ApJ...475..479W}
{White}, R.~L., {Becker}, R.~H., {Helfand}, D.~J., \& {Gregg}, M.~D. 1997,
  \href{http://dx.doi.org/10.1086/303564}{\JournalTitle{\apj}, 475, 479}

\bibitem[{{Wilhite} {et~al.}(2008){Wilhite}, {Brunner}, {Grier}, {Schneider},
  \& {vanden Berk}}]{wilhite2008variability}
{Wilhite}, B.~C., {Brunner}, R.~J., {Grier}, C.~J., {Schneider}, D.~P., \&
  {vanden Berk}, D.~E. 2008,
  \href{http://dx.doi.org/10.1111/j.1365-2966.2007.12655.x}{\JournalTitle{\mnras},
  383, 1232}

\bibitem[{{Wilhite} {et~al.}(2005){Wilhite}, {Vanden Berk}, {Kron},
  {Schneider}, {Pereyra}, {Brunner}, {Richards}, \& {Brinkmann}}]{Wilhite2005}
{Wilhite}, B.~C., {Vanden Berk}, D.~E., {Kron}, R.~G., {et~al.} 2005,
  \href{http://dx.doi.org/10.1086/430821}{\JournalTitle{\apj}, 633, 638}

\bibitem[{{Wilson} \& {Colbert}(1995)}]{Wilson1995}
{Wilson}, A.~S., \& {Colbert}, E.~J.~M. 1995,
  \href{http://dx.doi.org/10.1086/175054}{\JournalTitle{\apj}, 438, 62}

\bibitem[{{Wold} {et~al.}(2007){Wold}, {Brotherton}, \& {Shang}}]{Wold2007}
{Wold}, M., {Brotherton}, M.~S., \& {Shang}, Z. 2007,
  \href{http://dx.doi.org/10.1111/j.1365-2966.2006.11364.x}{\JournalTitle{\mnras},
  375, 989}

\bibitem[{{Wu} \& {Shen}(2022)}]{Wu2022}
{Wu}, Q., \& {Shen}, Y. 2022,
  \href{http://dx.doi.org/10.3847/1538-4365/ac9ead}{\JournalTitle{\apjs}, 263,
  42}

\bibitem[{{Yang} {et~al.}(2018){Yang}, {Wu}, {Fan}, {Jiang}, {McGreer},
  {Shangguan}, {Yao}, {Wang}, {Joshi}, {Green}, {Wang}, {Feng}, {Fu}, {Yang},
  \& {Liu}}]{Yang2018}
{Yang}, Q., {Wu}, X.-B., {Fan}, X., {et~al.} 2018,
  \href{http://dx.doi.org/10.3847/1538-4357/aaca3a}{\JournalTitle{\apj}, 862,
  109}

\bibitem[{{Zamaninasab} {et~al.}(2014){Zamaninasab}, {Clausen-Brown},
  {Savolainen}, \& {Tchekhovskoy}}]{Zamaninasab2014}
{Zamaninasab}, M., {Clausen-Brown}, E., {Savolainen}, T., \& {Tchekhovskoy}, A.
  2014, \href{http://dx.doi.org/10.1038/nature13399}{\JournalTitle{\nat}, 510,
  126}

\bibitem[{{Zheng} {et~al.}(2011){Zheng}, {Yuan}, {Gu}, \& {Lu}}]{Zheng2011}
{Zheng}, S.-M., {Yuan}, F., {Gu}, W.-M., \& {Lu}, J.-F. 2011,
  \href{http://dx.doi.org/10.1088/0004-637X/732/1/52}{\JournalTitle{\apj}, 732,
  52}

\bibitem[{{Zuo} {et~al.}(2012){Zuo}, {Wu}, {Liu}, \& {Jiao}}]{Zuo2012}
{Zuo}, W., {Wu}, X.-B., {Liu}, Y.-Q., \& {Jiao}, C.-L. 2012,
  \href{http://dx.doi.org/10.1088/0004-637X/758/2/104}{\JournalTitle{\apj},
  758, 104}

\end{thebibliography}

\end{document}